\documentclass[twocolumn,pre,floats,aps,amsmath,amssymb]{revtex4}

\usepackage{graphicx}
\usepackage{bm}
\usepackage{braket}
\usepackage{amsmath}
\usepackage{color}
\usepackage{subfigure}

\newcommand{\inst}{\hbox{\scriptsize inst}}
\newcommand{\eff}{\text{eff}}
\newcommand{\new}{\prime}
\newcommand{\ct}{\hbox{\scriptsize ct}}
\newcommand{\rel}{\text{rel}}
\newcommand{\tot}{\text{tot}}

\begin{document}
\title{Basis light-front quantization approach to positronium}

\author{Paul Wiecki, Yang Li, Xingbo Zhao, Pieter Maris and James P. Vary}
\affiliation{Department of Physics and Astronomy, Iowa State University, Ames, Iowa  50011  USA}
\date{\today}

\begin{abstract}
We present the first application of the recently developed Basis Light-Front Quantization (BLFQ) method to self-bound systems 
in quantum field theory, using the positronium system as a test case. Within the BLFQ framework, we develop a two-body 
effective interaction, operating only in the lowest Fock sector, that implements photon exchange, neglecting
fermion self-energy effects. 
We then solve for the mass spectrum of this interaction at the unphysical coupling $\alpha=0.3$. The resulting spectrum is in good agreement with
the expected Bohr spectrum of non-relativistic quantum mechanics. We examine in detail the dependence of the results on the 
regulators of the theory.

\end{abstract}

\maketitle

\section{Introduction}
\label{sec:intro}
The {\it ab initio} calculation of hadron mass spectra and observables in terms of their underlying quark and gluon
degrees of freedom remains a significant challenge to theoretical physics. 
Bound state problems in a quantum field theory such as QCD are inherently non-perturbative. 
Hadron problems are also many-body in nature and must be solved at strong coupling. 
Due to the complexity of the problem, seemingly simple quantities, such as 
the proton magnetic moment, have not yet been accurately calculated.

Observables of interest include the mass spectrum of the hadronic system (in particular  
``exotic'' states with quantum numbers beyond the constituent quark model), along with 
properties of the corresponding eigenstates. 
These include transition and decay rates, electric and magnetic moments, form factors,
structure functions and generalized parton distributions (GPDs).

Basis Light-Front Quantization (BLFQ) \cite{vary} is a promising tool for 
tackling hadron problems from first principles. BLFQ is a Hamiltonian-based approach that 
combines the advantages of light-front dynamics \cite{brodskyreview,harinotes}
with modern developments in {\it ab initio} nuclear structure calculations, such as the
No-Core Shell Model (NCSM) \cite{ncsm,Navratil:2000gs}. 
The similarity of light-front Hamiltonian quantum field theory to non-relativistic quantum many-body theory 
allows the quantum 
field theoretical bound state problem to be formulated as a large, sparse matrix diagonalization problem. State-of-the-art methods developed for
NCSM calculations can then be used to address hadronic systems \cite{methods1,methods2,methods3}.
The diagonalization of the light-front Hamiltonian in a Fock-space basis yields the mass eigenstates of the system, along with amplitudes for evaluating 
non-perturbative observables.

Recent works have successfully applied BLFQ to the single-electron
problem in QED in order to evaluate the electron 
anomalous magnetic moment both with \cite{heliprl}  
and without \cite{Zhao:2011ct,Zhao:2014xaa}
an external trap.  
Another recent application evaluates the electron GPDs \cite{Chakrabarti:2014cwa}.  
In addition, BLFQ has been extended 
to time-dependent strong external field problems (tBLFQ) 
such as non-linear Compton scattering \cite{xingboanton,Zhao:2013jia}.

Here, we investigate the positronium system as a test case for applying BLFQ to self-bound systems.
Our primary purpose is to confirm that BLFQ is capable of generating the expected Bohr spectrum of positronium, including relativistic effects
such as the hyperfine splitting of the ground state.

For this initial test case, we implement a two-body effective interaction that operates only in the lowest Fock sector. The effective interaction implements the
one-photon-exchange kernel necessary for Coulomb binding, but neglects the fermion self energy.
In future applications involving dynamical photons in the basis, a non-perturbative renormalization scheme will be needed to deal with fermion 
self-energy effects. 
We intend to use the results of this calculation as a benchmark for implementing a Fock-sector dependent renormalization scheme. 

We begin by introducing the basic elements of BLFQ, such as our choice of basis and truncation scheme. We then detail the derivation of the 
two-body effective interaction
and solve for the spectrum of positronium using this interaction. Finally, we examine in detail the dependence of the results on the regulators of our theory. 
Preliminary results were reported in Refs. \cite{krakow,wiecki}.

\section{Basis Light-Front Quantization}
\label{sec:blfq}
In principle, hadron observables can be evaluated by solving the eigenvalue equation
\begin{equation}
P^\mu P_\mu \ket{\Psi}=M^2 \ket{\Psi},
\label{eq:bound_st}
\end{equation}
where $P^\mu$ is the energy-momentum 4-vector operator. In BLFQ, we express the operator $P^2$ in 
light-cone gauge. The operator $P^2$ then plays the role of the 
Hamiltonian operator in non-relativisitic Quantum Mechanics. As such, it is sometimes referred as the
``light-cone Hamiltonian'' $H_{LC}\equiv P^2$. (Note that in this convention the ``Hamiltonian'' has energy squared
units.) This operator can be derived from any field theoretical Lagrangian via the Legendre transform. 
In BLFQ, Eq. \eqref{eq:bound_st} is expressed in a truncated basis, and the resulting finite-dimensional matrix is diagonalized numerically.
One then examines the trends in observables as the basis truncation is relaxed to estimate the results in the 
infinite matrix (or ``continuum'') limit.

Of course $H_{LC}$, being field theoretical in origin, contains terms which change particle number. Thus the basis space for 
performing a diagonalization must be expanded to include states with any number or species of particles. For example,
the positronium wavefunction could be expressed schematically as
\begin{align}
\Ket{e^+e^-}_{\hbox{\scriptsize phys}}&=a\Ket{e^+e^-}+b\Ket{e^+e^-\gamma}+c\Ket{e^+e^-\gamma\gamma}\nonumber\\
&+d\Ket{\gamma}+f\Ket{e^+e^-e^+e^-}+\cdots.
\end{align}
When $H_{LC}$ is derived from the QED or QCD Lagrangian, the resulting interactions 
change particle number by at most two. The resulting matrix will then be extremely sparse for a many-body calculation. 

\subsection{Basis and Truncation Scheme}
\label{sec:basis}
In order to numerically diagonalize $H_{LC}$, the infinite dimensional basis must be truncated down to a finite dimension. 
In BLFQ, three separate truncations are made.

First, the number of Fock sectors in the basis is truncated. This truncation will be based on physical as well as practical considerations.
For instance, positronium is expected to be fairly well described by the lowest few sectors. Thus, in this introductory work, we limit ourselves
to only the $\Ket{e^+e^-}$ and $\Ket{e^+e^-\gamma}$ sectors. We do not make any attempt here to examine the limit of increasing the 
number of Fock sectors.

Secondly, we discretize the longitudinal momentum by putting our
system in a longitudinal box of length $L$ and applying periodic boundary conditions (BCs). Specifically, we choose
periodic BCs for bosons and anti-periodic BCs for fermions. Thus
\begin{equation}
p^+=\frac{2\pi}{L}j,
\end{equation}
where $j$ is an integer for bosons, or a half-integer for fermions. For bosons, we exclude the ``zero modes'', i.e. $j\neq0$. 
In the many-body basis, all basis states are selected to have the same total longitudinal momentum $P^+=\sum_ip_i^+$,
where the sum is over the particles in a particular basis state. We then parameterize $P^+$ using a dimensionless variable $K=\sum_i j_i$ such that 
$P^+=\frac{2\pi}{L}K$. For a given particle $i$, the longitudinal momentum fraction $x$ is defined as
\begin{equation}
x_i=\frac{p_i^+}{P^+}=\frac{j_i}{K}.
\end{equation}
Due to the positivity of longitudinal momenta on the light-front \cite{harinotes}, fixing $K$ also serves as
a Fock space cutoff and makes the number of longitudinal modes finite \cite{1+1}. It is easy to see that $K$
determines our ``resolution'' in the longitudinal direction, and thus our resolution on parton distribution functions.
The longitudinal continuum limit corresponds to the limit $L,K \to \infty$.

Finally, in the transverse direction, we employ a 2D Harmonic Oscillator (HO) basis.
On the light-front, the generating operator for the 2D HO basis can be expressed as
\begin{align}
P_+^\Omega&=\sum_i\left(\frac{\mathbf{p}_i^2}{2p_i^+}+\frac{\Omega^2}{2}p_i^+\mathbf{r}_i^2   \right)\nonumber\\
&=\frac{\Omega}{2}\sum_i\left(\frac{\mathbf{p}_i^2}{x_iP^+\Omega}+x_iP^+\Omega \mathbf{r}_i^2   \right).
\label{eq:gen1}
\end{align}
Here, and elsewhere, the boldface type is reserved for 2D transverse vectors.
Each value of $\Omega$ in \eqref{eq:gen1} determines a unique complete and orthonormal basis.
By defining the new coordinates \cite{krakow}
\begin{align}
&\mathbf{q}\equiv\frac{\mathbf{p}}{\sqrt{x}}\nonumber,\\
&\mathbf{s}\equiv\sqrt{x}\mathbf{r},
\label{eq:mariscoords}
\end{align}
we can write the generating operator as
\begin{equation}
P_+^\Omega=\frac{\Omega}{2}\sum_i\left[\left(\frac{\mathbf{q}_i}{\sqrt{P^+\Omega}}\right)^2+\left(\sqrt{P^+\Omega}\,\mathbf{s}_i\right)^2   \right].
\label{eq:gen_op}
\end{equation}
We see that $P_+^\Omega$ generates a basis of energy scale $b=\sqrt{P^+\Omega}$.
In this work, we use Eq. \eqref{eq:gen_op} to define the 2D HO basis states.
The momentum-space eigenfunctions of Eq. \eqref{eq:gen_op} are
\begin{equation}
\Psi_n^m(\mathbf{q})=\frac{1}{b}\sqrt{\frac{4\pi\times n!}{(n+|m|)!}}e^{im\phi}\rho^{|m|}e^{-\rho^2/2}L_n^{|m|}(\rho^2),
\end{equation}
where $\rho\equiv\frac{|\mathbf{q}|}{b}$ and $\phi=\arg(\mathbf{q})$. $L_n^m(x)$ is the generalized (or ``associated'') Laguerre polynomial. 

The basis is made finite by restricting the number of allowed oscillator quanta in each many-body basis state according to
\begin{equation}
\sum_i\left(2n_i+|m_i|+1\right)\leq N_{\max}.
\end{equation}
The transverse continuum limit corresponds to $N_{\max}\to\infty$. In addition, we use an ``M-scheme'' basis. That is, our many
body states have well defined values of the total angular momentum projection
\begin{equation}
M_J=\sum_i\left(m_i+s_i\right),
\end{equation}
where $s=\pm\frac{1}{2}$ is the fermion spin, but they do not have a well-defined total angular momentum $J$.

The choice of the 2D HO basis for BLFQ was made with the hadrons in mind. The HO potential
is a confining potential, and therefore its wavefunctions should form an ideal basis for systems subject to QCD confinement.
However, it is well known that the HO basis is not ideal for the Coulomb problem, due to the mismatch of the 
asymptotic behaviors of the wavefunctions (see, for example, Ref. \cite{sturmian}):
the HO wavefunctions have Gaussian tails, while the hydrogen wavefunctions have a slower exponential decay.
We therefore expect slow convergence in the positronium problem.

\subsection{Center-of-Mass Factorization}
\label{sec:cm}
In BLFQ, we construct our many-body basis in single-particle coordinates. The rationale for doing this is its straightforward
generalization to a basis of many particles. In principle, relative (Jacobi) coordinates could be used, but this process rapidly 
becomes intractable as the particle number is increased, due to the need for proper symmetrization of the basis states. 
Of course, for the two-particle positronium system Jacobi coordinates
would be ideal. However, viewing the positronium problem only as a test case for our larger framework, here we 
nonetheless choose to work in single-particle coordinates.

Since our basis is constructed in single-particle coordinates, the center-of-mass (CM) motion of the system
is contained in our solutions. The use of the HO
basis combined with the $N_{\max}$ truncation is a great advantage here since it allows for the exact factorization of the wavefunction into ``intrinsic'' 
and ``CM'' components, even within a truncated basis. The CM motion can then be removed from the low-lying spectrum by introducing a 
Lagrange multiplier (also known as the Lawson term) to the Hamiltonian \cite{caprio}. The extra term
essentially makes CM excitations very costly energetically therefore removing spurious CM excitations from the low-lying spectrum.

When the Hamiltonian is expressed in terms of the coordinates \eqref{eq:mariscoords} exact CM factorization is achieved for all eigenstates,
even in a basis with arbitrary numbers of sectors, which is the reason for the introduction of these coordinates \cite{krakow}.
The CM motion is then governed by
\begin{equation}
H_{CM}=\left(\sum_{i}\sqrt{x_i}\mathbf{q}_i\right)^2+b^4\left(\sum_{i}\sqrt{x_i}\mathbf{s}_i\right)^2.
\end{equation}
The CM motion can be removed from the low-lying spectrum by adding a Lagrange multiplier proportional to $H_{CM}$ to the Hamiltonian to get
\begin{equation}
H'=H+\lambda\left(H_{CM}-2b^2I\right),
\end{equation}
where $H\equiv H_{LC}$. 
In practice, one selects $\lambda$ to be large enough that $2\lambda b^2$ is well above the excitation spectrum of interest. 
Demonstrations of the exact CM factorization within BLFQ are given in Refs. \cite{krakow,yang}.

\section{The Hamiltonian}
\label{sec:veff}

\subsection{Basic Structure}
We truncate the Fock space to include only $\Ket{e^+e^-}$ and $\Ket{e^+e^-\gamma}$ states. We wish to 
formulate an effective potential acting only in the $\Ket{e^+e^-}$ space that includes the effects 
generated by the $\Ket{e^+e^-\gamma}$ space. In the formalism of effective potentials, we consider the
$P$ space to be the $\Ket{e^+e^-}$ space and $Q$ space to be the $\Ket{e^+e^-\gamma}$ space. Let ${\cal P}$ 
be the operator that projects onto the $P$ space, and ${\cal Q }$ be the operator that projects onto the $Q$ space.

In addition to the effective interaction, the complete Hamiltonian will also include those terms from $H=H_{LC}$ which act directly in
the two-particle space: ${\cal P}H{\cal P}$. This contains two pieces. First, it contains the
two-particle kinetic energy. Secondly, it contains the light-front instantaneous photon exchange interaction. Thus the
total Hamiltonian can be expressed as
\begin{equation}
\Bra{f}{\cal P}H_{\tot}{\cal P}\Ket{i}=\Bra{f}{\cal P}\left(H_0+H_{\inst}+H_{\eff}\right){\cal P}\Ket{i},
\end{equation}
where states $\ket{i}$ and $\ket{f}$ are states in $P$ space ($\Ket{e^+e^-}$).
The basis states $\ket{i}$ and $\ket{f}$ are eigenstates of the free Hamiltonian (i.e. $H_0\ket{i}=\epsilon_i\ket{i}$) with eigenvalue
\begin{equation}
\epsilon_i=\sum_j\frac{\mathbf{p}_j^2+m_j^2}{x_j},
\label{h0}
\end{equation}
where the sum runs over particles (of mass $m_j$) in the state $\ket{i}$.

\subsection{Two-Body Effective Interaction}
We choose the Bloch form of the effective Hamiltonian.
The Bloch Hamiltonian \cite{dipankar} is given by:
\begin{align}
\Bra{f}H_{\eff}\Ket{i}&=\frac{1}{2}\sum_{n}\Bra{f}{\cal P}H{\cal Q}\Ket{n}\Bra{n}{\cal Q }H{\cal P}\Ket{i}\nonumber\\
&\times\left[\frac{1}{\epsilon_i-\epsilon_n}+
\frac{1}{\epsilon_f-\epsilon_n}\right].
\label{eq:bloch}
\end{align}
Here, $H=H_{LC}=P^2$ is the light-cone Hamiltonian introduced above.
States $\ket{i}$ and $\ket{f}$ are states in $P$ space ($\Ket{e^+e^-}$), while state $\ket{n}$ is in the $Q$ space ($\Ket{e^+e^-\gamma}$).
We have written Eq. \eqref{eq:bloch} in the notation traditionally used for effective potentials in non-relativistic 
Quantum Mechanics calculations. One must remember, though, that both the ``Hamiltonian'' $H$ and the ``unperturbed energy'' $\epsilon$
have energy squared units in our formalism.
Note that if $\ket{i}=\ket{f}$ this reduces to the usual formula from second-order
energy shift in perturbation theory. The derivation of \eqref{eq:bloch}, based on a perturbative expansion of the Okubo-Lee-Suzuki
effective Hamiltonian \cite{leesuzuki1,leesuzuki2,leesuzuki3,leesuzuki4,leesuzuki5,leesuzuki6}, is given in Ref. \cite{dipankar}. 

Since we are interested in primarily the effects of repeated photon exchange, we will only include those combinations of 
terms in ${\cal P}H{\cal Q}$ and ${\cal Q}H{\cal P}$ which generate the photon exchange. We neglect the combinations 
which result in the photon being emitted and absorbed by the same fermion. That is, we do not incorporate the fermion self-energy.
In addition, we work with unit-normalized eigenstates and a fixed value of the coupling constant. Details are presented in the Appendix.

\subsubsection{Light-front small-$x$ singularities}
The instantaneous photon exchange interaction $H_{\inst}$ contains a singularity of the form $\frac{1}{(x_1-x_1')^2}$, where $x_1$ ($x_1'$) 
is the longitudinal momentum fraction of the incoming (outgoing) fermion (see Eq. \eqref{eq:inst}).
In light-front $S$-matrix perturbation theory, the amplitude for electron-positron scattering via a dynamical photon
contains a term identical to the instantaneous photon exchange interaction, but with opposite sign. Thus the instantaneous 
interaction is cancelled {\it in its entirety}, leaving behind the familiar equal-time Feynman amplitude \cite{harinotes,brodskyreview}. 
From this perspective, the instantaneous photon exchange interaction exists only to cancel this singularity in the 
light-front electron-positron scattering amplitude. This singularity is an artifact of
the use of light-cone gauge, and must be cancelled.

In our non-perturbative calculation, this cancellation of small-$x$ singularities does not occur in general.
To remove the unphysical singularity in $H_{\inst}$, we introduce a counterterm of the form
\begin{align}
\Bra{f}H_{\ct}\Ket{i}&=-\sum_{n}\Bra{f}{\cal P}H{\cal Q}\Ket{n}\Bra{n}{\cal Q}H{\cal P}\Ket{i}\nonumber\\
&\times\left[\frac{\left(a-b\right)^2}{2ab\left(a+b)\right)}\right],
\label{eq:ct}
\end{align}
where $a=\epsilon_i-\epsilon_n$ and $b=\epsilon_f-\epsilon_n$. The resulting effective potential, $H^{\new}_{\eff}$, is 
\begin{eqnarray}
\Bra{f}H^{\new}_{\eff}\Ket{i}&=&\Bra{f}\left(H_{\eff}+H_{\ct}\right)\Ket{i}
\nonumber \\
&=&
\sum_{n}\frac{\Bra{f}{\cal P}H{\cal Q}\Ket{n}\Bra{n}{\cal Q}H{\cal P}\Ket{i}}{\frac{1}{2}
\left[\left(\epsilon_i-\epsilon_n\right)+\left(\epsilon_f-\epsilon_n\right)\right]}.
\end{eqnarray}
In this form the cancellation of the instantaneous diagram does occur, and $H_{\inst}+H^{\new}_{\eff}$ is free of unphysical
light-front small-$x$ singularities. Details are given in the Appendix.
We note that our choice of counterterm is equivalent to the prescription used
in previous work in Discretized Light-Cone Quantization (DLCQ) \cite{pauliwolz,trittmanpauli,muonium}. 
Refs. \cite{pauliwolz,trittmanpauli} also provide arguments for the plausibility of this prescription.

\subsubsection{Final Expression}
Our Hamiltonian is then
\begin{equation}
\Bra{f}{\cal P}H_{\tot}{\cal P}\Ket{i}=\Bra{f}{\cal P}\left(H_0+H_{\inst}+H_{\eff}^{\new}\right){\cal P}\Ket{i},
\end{equation}
where, after canceling the instantaneous interaction, we are left with ($\alpha=g^2/4\pi$)
\begin{align}
&H_{\inst}+H_{\eff}^{\new}=\frac{\alpha}{K}\sum_{\bar{\alpha}_1\bar{\alpha}_1'\bar{\alpha}_2\bar{\alpha}_2'}\delta^{j_1'+j_2'}_{j_1+j_2}
b_{\bar{\alpha}_1'}^\dagger d^\dagger_{\bar{\alpha}_2'}d_{\bar{\alpha}_2}b_{\bar{\alpha}_1}\nonumber\\
&\times\sqrt{x_1x_2x_1'x_2'}
\int \frac{d^2\mathbf{q}_1}{(2\pi)^2} \frac{d^2\mathbf{q}_1'}{(2\pi)^2} \frac{d^2\mathbf{q}_2}{(2\pi)^2} \frac{d^2\mathbf{q}_2'}{(2\pi)^2}\nonumber \\
&\times\frac{(2\pi)^2\delta^{(2)}(\sqrt{x_1}\mathbf{q}_1+\sqrt{x_2}\mathbf{q}_2-\sqrt{x_1'}\mathbf{q}_1'-\sqrt{x_2'}\mathbf{q}_2')}
{(x_1-x_1')\frac{1}{2}\left[\left(\epsilon_i-\epsilon_n\right)+\left(\epsilon_f-\epsilon_n\right)\right]}\nonumber\\
&\times\Psi_{n_1}^{m_1}(\mathbf{q}_1)\Psi_{n_2}^{m_2}(\mathbf{q}_2)\Psi_{n_1'}^{m_1'\ast}(\mathbf{q}_1')\Psi_{n_2'}^{m_2'\ast}(\mathbf{q}_2')\nonumber \\
&\times S_{\alpha_1,\alpha_2,\alpha_1',\alpha_2'}(\sqrt{x_1}\mathbf{q}_1,\sqrt{x_2}\mathbf{q}_2,\sqrt{x_1'}\mathbf{q}_1',\sqrt{x_2'}\mathbf{q}_2'),
\end{align}
with $\bar{\alpha}_i$ representing the set of discrete quantum numbers $(j_i,s_i,n_i,m_i)$ and 
$\alpha_i$ representing the subset $(j_i,s_i)$. If the fermions have mass $m_f$ and the photon has mass $\mu$,
the parts of the energy denominator are given by
\begin{align}
\epsilon_i-\epsilon_n&=
\frac{(\sqrt{x_1}\mathbf{q}_1)^2+m_f^2}{x_1}-\frac{(\sqrt{x_1'}\mathbf{q}'_1)^2+m_f^2}{x_1'}\nonumber\\
&-\frac{(\sqrt{x_1}\mathbf{q}_1-\sqrt{x_1'}\mathbf{q}_1')^2+\mu^2}{x_1-x_{1'}}\nonumber\\
-(\epsilon_f-\epsilon_n)&=
\frac{(\sqrt{x_2}\mathbf{q}_2)^2+m_f^2}{x_2}-\frac{(\sqrt{x_2'}\mathbf{q}'_2)^2+m_f^2}{x_2'}\nonumber\\
&-\frac{(\sqrt{x_2}\mathbf{q}_2-\sqrt{x_2'}\mathbf{q}_2')^2+\mu^2}{x_2-x_{2'}}.
\end{align}
The explicit expression for the spinor part, 
$S_{\alpha_1,\alpha_2,\alpha_1',\alpha_2'}(\sqrt{x_1}\mathbf{q}_1,\sqrt{x_2}\mathbf{q}_2,\sqrt{x_1'}\mathbf{q}_1',\sqrt{x_2'}\mathbf{q}_2')$
is provided in the Appendix (Table \ref{tab:spinor}). 
The highly oscillatory 8D integration can be evaluated using repeated 2D Talmi-Moshinsky transformations \cite{Talmi}. 
The integral can be reduced down to a single 2D integral, which is evaluated numerically. The details are presented in the Appendix.

Note, also, that a fictitious photon mass $\mu$ has been introduced to regulate the expected Coulomb singularity that, while integrable,
introduces numerical difficulties. We will later examine the physical limit $\mu\to 0$.

\subsection{Regulated Effective Interaction}
\label{sec:mod}
Previous authors investigating the problem of positronium on the light-front with a one-photon exchange kernel have noted a very small
dependence of the ground state energy on the ultraviolet cutoff of the theory, corresponding to a logarithmic 
divergence  \cite{pauliwolz,karmanov,muonium}. 
Both Refs. \cite{pauliwolz,karmanov}
state that the origin of the instability can be traced to a 
particular term in the effective interaction (or one-photon exchange kernel) which tends to a non-zero constant, in momentum space, at
asymptotically large momentum transfer, corresponding to a Dirac delta potential in coordinate space. Since the 2D Dirac delta potential well
has no bound states of finite binding energy \cite{deltawell}, this leads to a divergence.

The same divergent term is indeed present in our effective interaction. Numerically, we find that the ground state energy 
is unstable with increasing $N_{\max}$, but stable with respect to $K$ for a fixed $N_{\max}$.
The divergence can be removed by introducing a counterterm. 
Ref. \cite{pauliwolz} argues that such a counterterm can be understood as accounting for the effects of the
$\Ket{e^+e^-\gamma\gamma}$ Fock sector.
The necessary counterterm can be expressed as a modification of the 
spinor part of the Hamiltonian,
$S_{\alpha_1,\alpha_2,\alpha_1',\alpha_2'}(\sqrt{x_1}\mathbf{q}_1,\sqrt{x_2}\mathbf{q}_2,\sqrt{x_1'}\mathbf{q}_1',\sqrt{x_2'}\mathbf{q}_2')$,
according to
\begin{equation}
S\to S- 2\left[\mathbf{q}_{\rel}^2\!+\!\mathbf{q}_{\rel}'^2\right]
\delta_{s_1}^{s_1'}\delta_{s_2}^{s_2'}(\delta_{s_1}^+\delta_{s_2}^-+\delta_{s_1}^-\delta_{s_2}^+),
\label{eq:modified}
\end{equation}
where $s_i=\pm$ represents the fermion spin.
In single-particle coordinates, we have $\mathbf{q}_{\rel}=\sqrt{x_2}\mathbf{q}_1-\sqrt{x_1}\mathbf{q}_2$ and
$\mathbf{q}_{\rel}'=\sqrt{x_2'}\mathbf{q}_1'-\sqrt{x_1'}\mathbf{q}_2'$. We will refer to the resulting interaction as the 
``regulated'' effective interaction. We use the term ``unregulated'' to refer to the original effective interaction.

\section{Numerical Results}
\label{sec:results}
In non-relativistic Quantum Mechanics, the hyperfine splitting between the $^1S_0$ and $^3S_1$ states of positronium scales as 
$\alpha^4$, where $\alpha$ is the fine structure constant. At physical coupling, the expected hyperfine splitting
and even the binding energy
are then uncomfortably small relative to the precision of our numerical integrals. Since we would like to use
the hyperfine splitting to test our BLFQ results, we use a large coupling of $\alpha=0.3$ to exaggerate both the
binding energy and the hyperfine splitting. We then compare our results not to experiment, but to the predictions
of non-relativistic Quantum Mechanics at this unphysical value of $\alpha$.
This value of $\alpha$ also allows a direct comparison to the DLCQ results of Ref. \cite{pauliwolz,muonium}.

The numerical results were obtained using the Hopper Cray XE6 and Edison Cray XC30 at NERSC. 
ScaLAPACK software \cite{scalapack} was used for the diagonalization.
In this particular implementation of BLFQ, the resulting matrix is quite dense. However, in future applications involving 
multiple Fock sectors, the matrix will be extremely sparse. 

We obtain exact CM factorization for both the regulated and unregulated interactions. The spurious CM motion is removed by 
using a Lagrange multiplier, as discussed in Sec. \ref{sec:cm}, so that all states shown below are in the ground state of CM motion.

\subsection{Spectrum}
\label{sec:spectrum}

\begin{figure}[t]
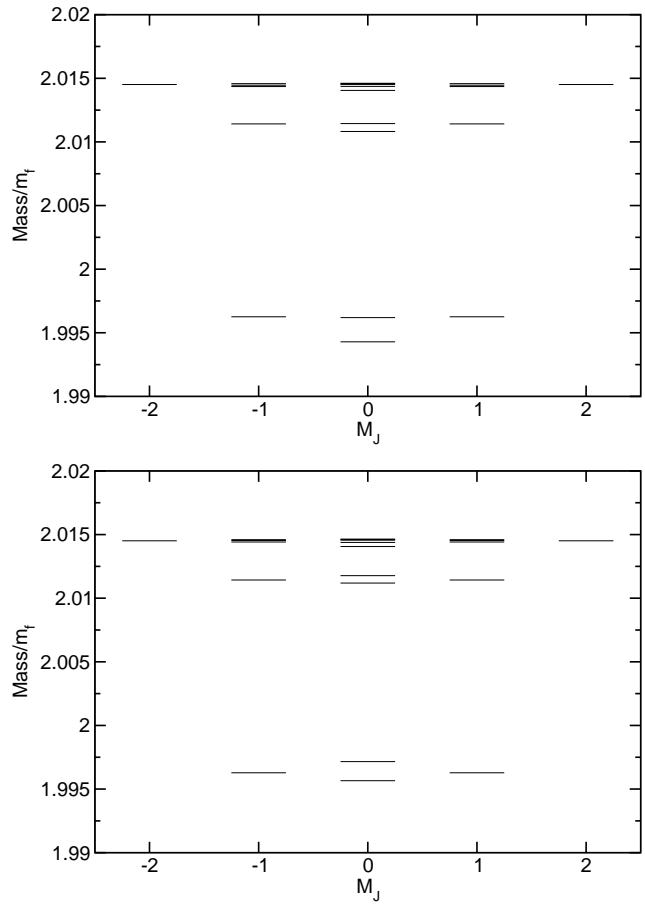

\centering
\subfigure{\includegraphics[width=3.3 in]{spectrum.eps}}
\quad
\subfigure{\includegraphics[width=3.3 in]{spectrum_mod.eps}}
\caption{Representative spectrum of positronium ($\alpha=0.3$) calculated in BLFQ at $K=N_{\max}=19$ and $\mu=0.1m_f$,
using the unregulated (top) and regulated (bottom) effective interactions. 
The exact energies shown should not be interpreted as final converged results. 
Using the unregulated interaction (top), the approximate
rotational invariance allows for the clear identification of the $1 ^1S_0$, $1 ^3S_1$, $2 ^1S_0$, $2 ^3S_1$, 
$2 ^1P_1$, $2 ^3P_0$, $2 ^3P_1$ and $2 ^3P_2$ states of the positronium system (see text for details). Using the regulated interaction
(bottom), the approximate rotational invariance is more strongly broken.}
\label{fig:spectrum}
\end{figure}

A representative spectrum is shown in Fig. \ref{fig:spectrum}. These results are produced with
$K=N_{\max}=19$ and $\mu=0.1m_f$. The energies shown are only representative and should not be considered
converged or final. The general features of the spectrum shown here are common to any calculation with $K=N_{\max}=19$ and above.
Convergence will be considered below. 

On the light-front, the total angular momentum operator is dynamical. 
In addition, manifest rotational invariance is lost in our calculation
due to Fock sector truncation, as well as the different discretizations in the longitudinal and transverse directions.
However, the total $J$ of the states can be extracted by examining the multiplet structure of the spectrum 
as it appears in the top panel of Fig. \ref{fig:spectrum}.
The ground state, for example, appears only in the $M_J=0$ calculation, suggesting that it has $J=0$.
We also see a triplet of states above the ground state with $M_J=-1,0,1$, suggesting that these states form a $J=1$ multiplet.
The lack of manifest rotational invariance is seen only in the lack of exact degeneracy between the states in 
this multiplet. The difference, however, is quite small, being approximately $1\%$ of the binding energy, 
and is nearly invisible on this scale. We therefore feel confident extracting $J$ by examining the number of 
states in each approximately degenerate multiplet.

In the non-relativistic quantum mechanics description of the positronium system, the states are labelled using the 
notation $n^{(2S+1)}L_J$, where $n$ is the principal quantum number of atomic physics, 
$S$ is the total intrinsic spin, $L$ is the total
orbital angular momentum (expressed in spectroscopic notation) and $J$ is the total angular momentum.
(Note that the $n$ of atomic physics is related to the 
radial node quantum number of nuclear physics by $n_{\text{atomic}}=n_{\text{radial}}+L+1$.)
The lowest two states are then $1 ^1S_0$ and $1 ^3S_1$, the singlet and triplet ground state configurations respectively.
The splitting between them is referred to as the hyperfine splitting. 
In a relativistic
theory such as BLFQ, $L$ and $S$ are no longer good quantum numbers. We identify the low-lying $J=0$ and $J=1$ multiplets in 
the BLFQ spectrum as the expected $1 ^1S_0$ and $1 ^3S_1$ states of positronium. For simplicity, we will 
label the states according to their non-relativistic description, but we stress that we can only extract 
$J$ (approximately) from our BLFQ calculation, and not $L$ or $S$.
Nevertheless, the BLFQ wavefunction of the state we identify as the $1 ^1S_0$ state is indeed anti-symmetric with respect to 
spin exchange, as one would expect for an $S=0$ state. (We reached this conclusion by examining the amplitude by eye at $N_{\max}=4$.)
Similarly, the $1 ^3S_1$ wavefunction is symmetric with respect to spin exchange, consistent with the non-relativistic
expectation. 

A similar grouping of states appear higher up in the spectrum. We identify these as the $2 ^1S_0$ and $2 ^3S_1$
states of positronium. The remaining four states can be identified, via similar reasoning, to be 
one $J=0$ multiplet, two $J=1$ multiplets, and one $J=2$ multiplet. (The highest line in the $M_J=0$ calculation is thicker
to indicate that it represents two nearly degenerate states.)
This is exactly what we expect for the $2P$ levels of 
positronium in non-relativistic quantum mechanics. We therefore identify these states with the expected
$2 ^1P_1$, $2 ^3P_0$, $2 ^3P_1$ and $2 ^3P_2$ positronium states. Note that in our BLFQ calculation, we cannot distinguish the two
$J=1$ states because we do not know $L$ and $S$. (In a relativistic notation, the states are distinguished by their charge conjugation
quantum number.)

In the lower panel of Fig. \ref{fig:spectrum}, the spectrum of the regulated interaction is shown.
The only difference is that the rotational invariance is more severely broken. Compared to 
the unregulated interaction, the $M_J=0$ states are shifted upwards, while the $M_J=\pm1$ states remain essentially unchanged.
A detailed investigation of the rotational symmetry breaking of the two interactions is beyond the scope of this paper.
For simplicity, we use the same state identifications for the regulated interaction as for the unregulated interaction.

\begin{figure}[t]
\centering
\includegraphics[width=3 in]{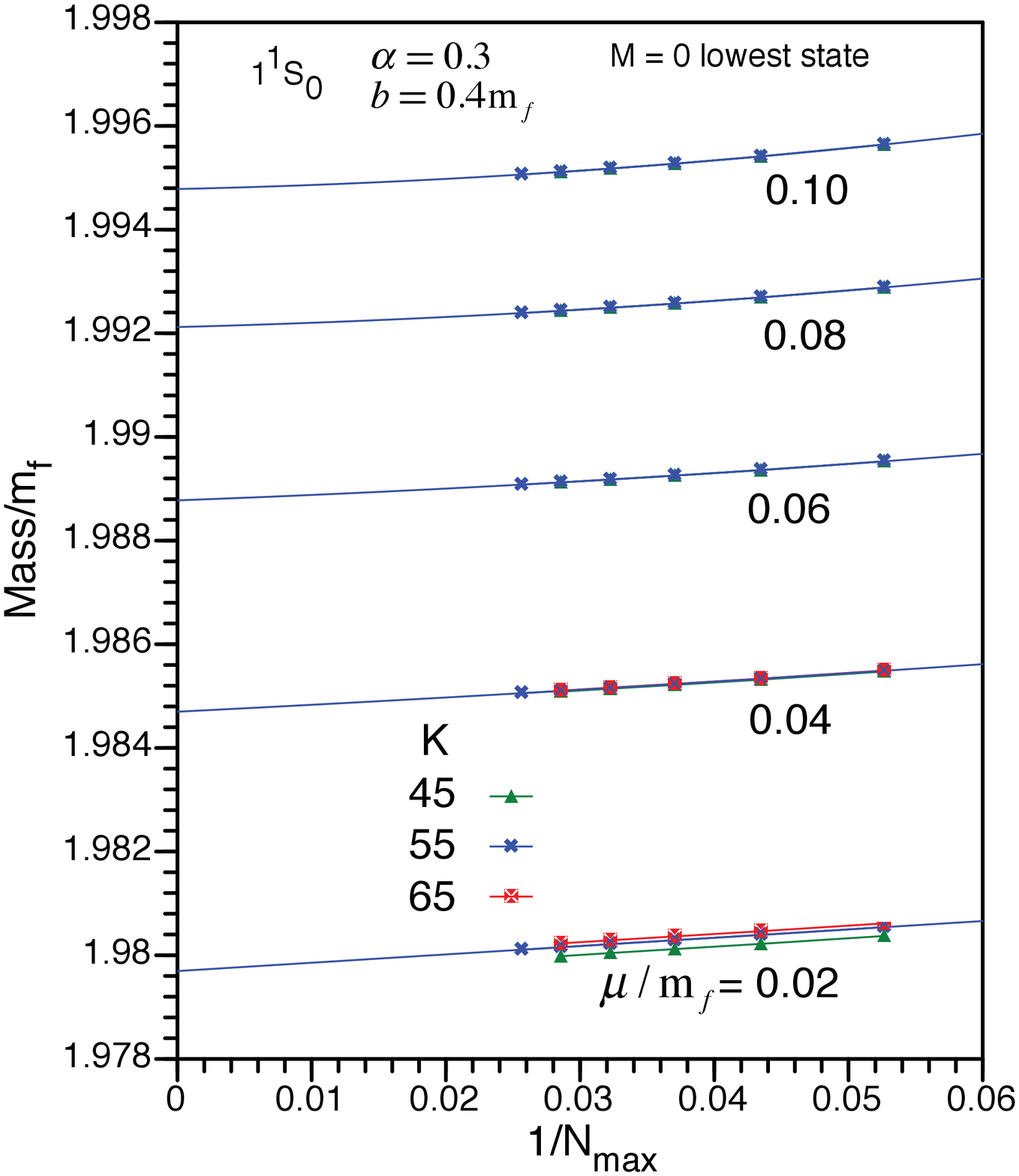}
\includegraphics[width=3 in]{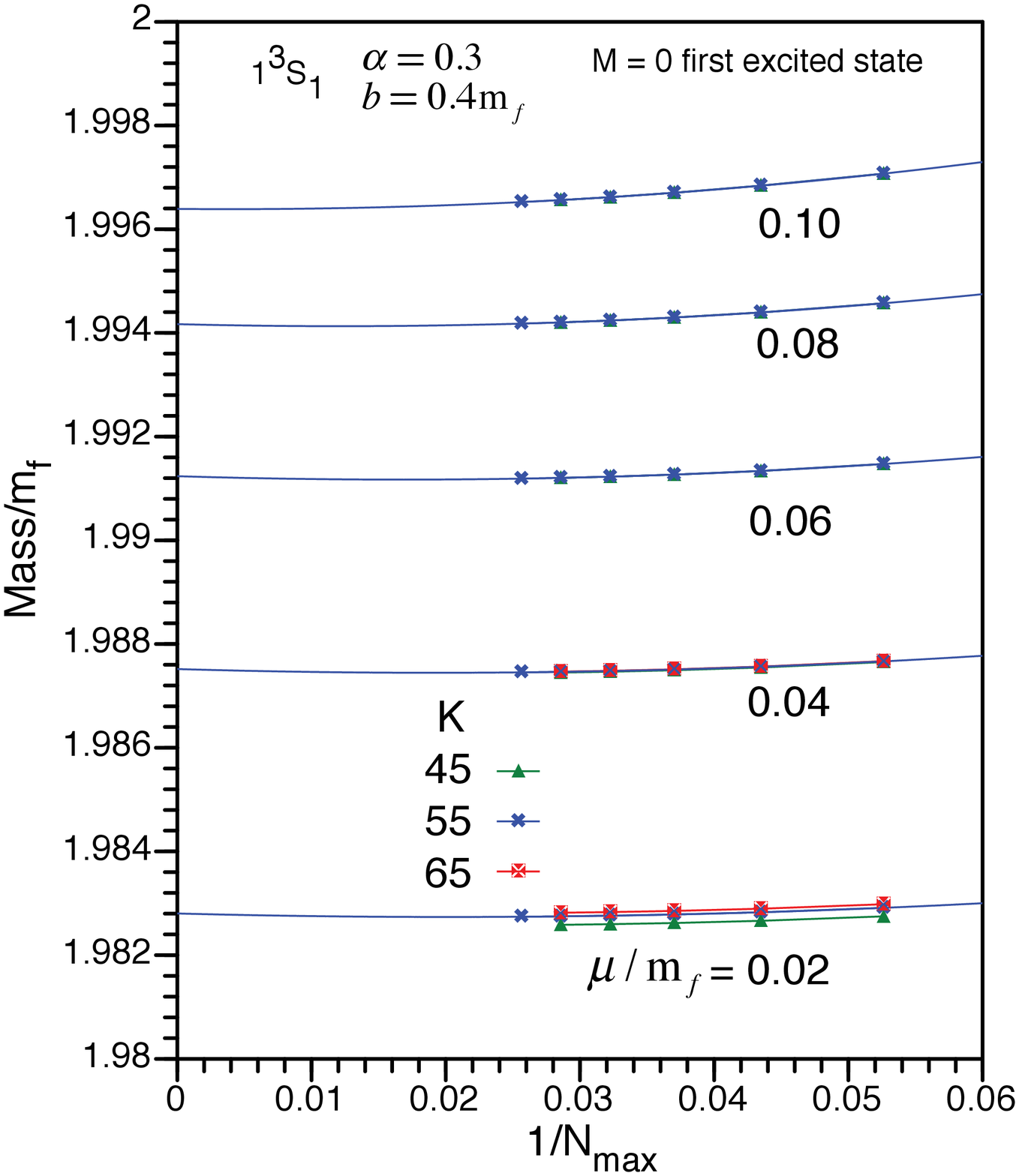}
\caption{(Color online) Singlet and triplet ground states of positronium ($\alpha=0.3$) as a function of $1/N_{\max}$ for 
indicated values of $K$ and $\mu$. 
The curves are second order polynomial fits of the $K=55$ results used to extrapolate the curves to the limit $N_{\max}\to\infty$ ($1/N_{\max}\to0$).
The basis energy scale $b=0.4m_f$ is chosen to optimize the $N_{\max}$ convergence rate for these states. }
\label{fig:ground}
\end{figure}

\subsection{Ground State and Hyperfine Splitting}
\label{sec:ground}

We now consider the dependence of the singlet and triplet ground states of positronium on the regulators of our theory.
In all of the results below, the value of the basis energy scale is taken to be $b=0.4m_f$. 
In principle, the results will converge for any value of $b$, since the basis is complete regardless of the energy scale. However,
in practice, the convergence rate is strongly affected by the value of $b$. To find the optimal value of the basis energy scale,
we plotted the ground state energy as a function of $b$ at $K=N_{\max}=25$, treating $b$ as a variational parameter. 
The curve was found to reach a minimum at $b=0.4m_f$. This value of the basis energy scale, then, is optimal for the convergence of 
the ground state energy with respect to $N_{\max}$. We emphasize, however, that a different state or observable may have a different 
optimal value. 

Throughout this work we calculate only at {\it odd} $K$. The reason is as follows. The kinetic energy term of $H_{LC}$ \eqref{h0}
has a term of the form $m_f^2\left(\frac{1}{x_1}+\frac{1}{x_2}\right)$. Since we require $x_1+x_2=1$, 
this term in the kinetic energy takes its minimum value when $x_1=x_2=\frac{1}{2}$. This leads to a binding threshold of $2m_f$ as expected.
If $K$ is odd, states with $x_1=x_2$ are present in the basis. If, however, $K$ is even, equal longitudinal momentum splitting is not present in
the basis and the ground state energy is increased by a term proportional to $m_f^2$. Since, in positronium, the fermion mass scale is much 
larger than the binding energy scale, the ground state energy is unnecessarily far from convergence if $K$ is even.

In Fig. \ref{fig:ground}, we plot the singlet and triplet ground state energies as a function
$1/N_{\max}$ for a series of values of the longitudinal resolution $K$ and fictitious photon mass $\mu$.
The two states are identified in our spectrum as the lowest two states of the $M_J=0$ calculation, as discussed in Sec. \ref{sec:spectrum}.
Since our Light-Cone Hamiltonian has energy squared units, the energy values plotted are the square root of the eigenvalues of the
Hamiltonian. 
We plot only the results using the regulated interaction, where convergence can be expected.
The range of values in $1/N_{\max}$ correspond to $N_{\max}\in[19,39]$.

The convergence in $K$ is rapid for $\mu\geq0.6m_f$, as one can see from the fact that the curves for $K=45$ and $K=55$ are nearly coincident.
As $\mu$ is decreased further, the $K$ convergence becomes slower, and we include the $K=65$ results for comparison.
Nevertheless, the results still display a clear converging trend with respect to $K$ for $\mu=0.02m_f$. 
The states also exhibit only mild dependence on $1/N_{\max}$, indicating convergence with respect to $N_{\max}$ also. 
To examine the continuum limit of $N_{\max}\to\infty$, we fit the $K=55$ curves to a second order polynomial 
in the variable $1/N_{\max}$ and extrapolate the fitted curve to the limit  $1/N_{\max}\to0$.

The dependence of the states on the fictitious photon mass $\mu$ will be discussed in more detail in Sec. \ref{sec:muto0}, where we 
examine the physical limit of $\mu\to0$ to compare our results with the standard non-relativistic quantum mechanics treatment.

\subsection{$2 ^3P_2$ State}
\label{sec:j=2}
As a representative excited state, we also consider the $2^3P_2$ state. Being the lowest $J=2$ state in the positronium spectrum, 
this state is identified in our spectrum as the lowest state of the $M_J=2$ calculation.
As discussed in Sec. \ref{sec:ground}, the optimal value of the H.O. basis energy scale, $b$, depends on the specific 
observable under examination. For the convergence of the ground state energy, we found an optimal value of $b=0.4m_f$.
However, this is not the optimal choice for the $2^3P_2$ state. The Coulomb well becomes much wider in position space
as one approaches the binding threshold. This indicates that the optimal H.O. basis length scale for convergence of the $2^3P_2$ state energy
will be larger than the optimal value for the ground state. A larger length scale, of course, corresponds to a smaller basis energy scale $b$.
Indeed, a variational calculation minimizing the $2^3P_2$ state energy, with $b$ as the variational parameter, 
indicates that $b=0.1m_f$ would give optimal convergence for this state. We therefore adopt this basis energy scale for our calculations
of the $2^3P_2$ state binding energy.

\begin{figure}[t]
\includegraphics[width=3.3 in]{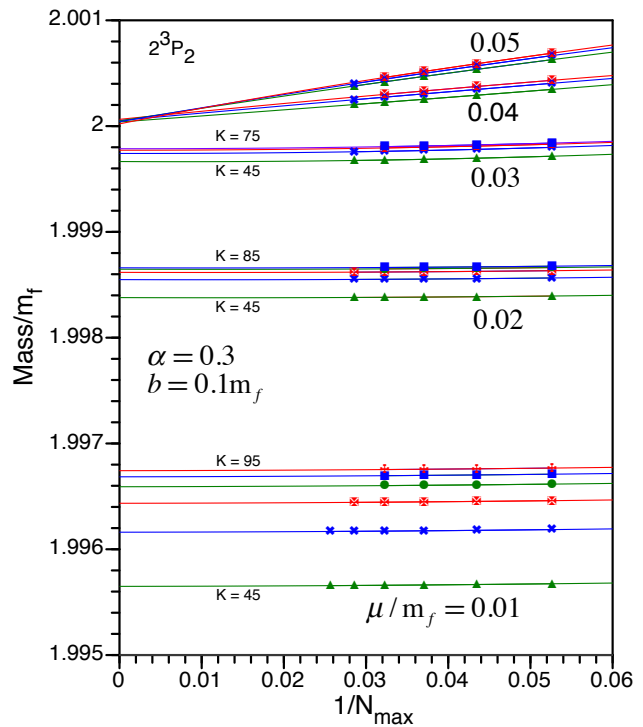}
\caption{(Color online) The $2 ^3P_2$ state of positronium ($\alpha=0.3$) as a function of $1/N_{\max}$ for indicated values of $K$ and $\mu$
(values of $K$ increase in increments of 10).
The discrete points represent the lowest mass eigenstate obtained in calculations with 
fixed total magnetic projection $M_J=2$.
The curves are second order polynomial fits used to extrapolate the curves to the limit $N_{\max}\to\infty$ ($1/N_{\max}\to0$).
The basis energy scale $b=0.1m_f$ is chosen to optimize the $N_{\max}$ convergence rate of the $2 ^3P_2$ state binding energy.
We include higher values of $K$ as $\mu$ is decreased because convergence with $K$ becomes slower. For $\mu\geq 0.04m_f$, 
the energy converges to $2m_f$ (the threshold for binding) in the continuum limit, indicating the 
the state is not bound at these high values of $\mu$.}
\label{fig:3P2}
\end{figure}

The dependence of the $2^3P_2$ state on $1/N_{\max}$ for a series of values of $K$ and $\mu$ is shown in Fig. \ref{fig:3P2}.
The convergence with respect to $K$ is similar to that of the ground state, becoming slower as $\mu$ is decreased towards zero.
We therefore include higher values of $K$ as $\mu$ is decreased to ensure a good estimate of the $K\to\infty$ limit. 
The trends in $1/N_{\max}$ are again fit to second order polynomials to examine the limit $N_{\max}\to\infty$ ($1/N_{\max}\to0$).
For $\mu\geq 0.04m_f$, the energy tends to $2m_f$ in the continuum limit, which is the threshold for binding. We conclude that, for $\mu\geq 0.04m_f$, 
the $2^3P_2$ state is a quasi-bound continuum state, and only becomes bound as the fictitious photon mass decreases below that value.

\subsection{$\mu\to0$ Limit}
\label{sec:muto0}
In this section, we compare our BLFQ results to the standard formulas of non-relativistic quantam mechanics. 
These predictions are (expressed in fermion mass units) \cite{bs}:
\begin{align}
M_{\text{singlet}}&=2-\frac{\alpha^2}{4}\left(1+\frac{63}{48}\alpha^2  \right) \label{eq:nrqm1}\\
M_{\text{triplet}}&=2-\frac{\alpha^2}{4}\left(1-\frac{1}{48}\alpha^2  \right) \label{eq:nrqm2}\\
M_{^3P_2}&=2-\frac{\alpha^2}{16}\left(1+\frac{43}{960}\alpha^2 \right) \label{eq:nrqm3}.
\end{align}
In these formulas, the $\alpha^2$ term corresponds to the Bohr energies and the $\alpha^4$ term incorporates 
first-order perturbative relativistic corrections, neglecting the possibility of a
temporary annihilation of the electron and positron into a virtual photon. Since we neglect the $\Ket{\gamma}$
Fock sector, our BLFQ results should be comparable to these predictions.

\begin{figure}[t]
\includegraphics[width=3.3 in]{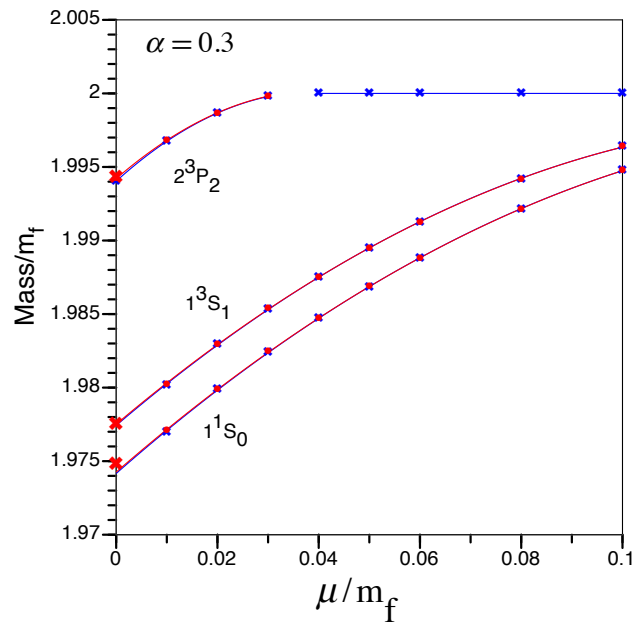}
\caption{(Color online) Dependence of the $1 ^1S_0$, $1 ^3S_1$ and $2 ^3P_2$ states on the fictitious photon mass $\mu$. The calculation
is performed at a large, unphysical coupling of $\alpha=0.3$.
The plotted blue crosses are the results of the $N_{\max}\to\infty$ extrapolations (at fixed $K$) performed in Figs. \ref{fig:ground} and \ref{fig:3P2}.
The fixed $K$ values chosen for the $1/N_{\max}$ extrapolations are 
$K=55$ for $\mu = 0.06 - 0.10$, $K=65$ for $\mu = 0.04 - 0.05$, $K=75$ for $\mu = 0.03$, $K=85$ for $\mu = 0.02$ and $K=95$ for $\mu = 0.01$.
These values of $K$ are chosen to be sufficiently high that the results can be considered converged with respect to $K$.  
For comparison, the results found using a simple $K\to\infty$ extrapolation (see text) are shown as solid red circles.
The curves are second order polynomial fits, used to extrapolate to the physical limit $\mu\to0$.
The crosses on the energy axis represent the predictions of non-relativistic quantum mechanics, including first-order 
perturbative relativistic corrections, as discussed in the text.}
\label{fig:muto0gs}
\end{figure}

In Fig. \ref{fig:muto0gs}, we plot the results of our (fixed $K$) $1/N_{\max}\to0$ extrapolations from Secs. \ref{sec:ground} and \ref{sec:j=2} 
as a function of the fictitious photon mass
regulator $\mu$. Recall that the convergence with respect to $K$ gets slower as $\mu$ is decreased. To account for this, we increase the
the fixed value of $K$ used in the $1/N_{\max}\to0$ extrapolations as $\mu$ gets smaller. For example, we use $K=55$ for $\mu\geq0.06m_f$, but 
$K=65$ for $\mu=0.04m_f$ and $\mu=0.05m_f$. We then use $K=75$ for $\mu=0.03m_f$, $K=85$ for $\mu=0.02m_f$ and $K=95$ for $\mu=0.01m_f$. 

We also make a simple extrapolation for the $K\to\infty$ limit at fixed $N_{\max}$ and fixed $\mu$.  This extrapolation was found by comparing 
many increments in such $K$ values corresponding to vertical sets of symbols in Figs. \ref{fig:ground} and \ref{fig:3P2} at fixed $\mu$ values.  
A stable estimator for the converged result is defined
by taking a result at $K$ and adding 1.25 times the difference between that result and the result at $K - 10$.  
That is, the estimators from several choices of K give a stable result to seven significant figures.  
The result of these extrapolations are indicated by the solid circles (red dots) in Fig. \ref{fig:muto0gs}.

We fit the resulting curves to second order polynomials in $\mu$ and extrapolate these curves to the 
physical limit of $\mu\to0$. For comparison, the predictions of Eqs. \eqref{eq:nrqm1}-\eqref{eq:nrqm3} are shown as crosses on the vertical axis.
The agreement is excellent, despite the large, non-physical value of $\alpha=0.3$. The ground state binding energy differs only by $5.84\times10^{-4}m_f$,
or $2.3\%$ of the binding energy. 
The hyperfine splitting in BLFQ, $3.26\times10^{-3}m_f$, is slightly larger than the expected value of $2.70\times10^{-3}m_f$.  
The $2 ^3P_2$ state binding energy is well-reproduced also.

\section{Conclusions and Outlook}
\label{sec:conclusion}
We have presented here the first application of BLFQ to self-bound systems in quantum field theory, using positronium
as the test case. In this work, we have truncated the Fock space to include only the $\Ket{e^+e^-}$ and $\Ket{e^+e^-\gamma}$
sectors. We further restricted the basis to the $\Ket{e^+e^-}$ sector alone by developing a two-body effective interaction,
incorporating the photon exchange effects generated by the $\Ket{e^+e^-\gamma}$ sector.
Diagonalization of the light-cone Hamiltonian including this interaction results in a repeated iteration of the effective interaction,
giving a non-perturbative solution to the bound state problem (equivalent to  
light-front ladder truncation with one-photon exchange kernel).
In this initial work, we have neglected fermion-self 
energy effects arising from the $\Ket{e^+e^-\gamma}$ sector.

Positronium is a particularly challenging test case for BLFQ using the HO basis due to the mismatch of the asymptotics of the 
basis states with the asymptotic bound state amplitudes of Coulombic systems, as discussed in Sec. \ref{sec:basis}.
Nevertheless, we have demonstrated how this mismatch may be overcome with extrapolations to obtain accurate results for the positronium spectrum.

Our converged BLFQ results are in good agreement with the Bohr spectrum of the positronium system at $\alpha=0.3$.
All of the expected total-$J$ multiplets arise in our calculation. In particular, the binding energies (including relativistic corrections)
of the $1 ^1S_0$, $1 ^3S_1$ and $2 ^3P_2$ states are reproduced quantitatively in the BLFQ continuum limit.
Thus these results confirm that BLFQ is capable of generating the expected spectrum for positronium. This 
calculation serves as a strong benchmark test for BLFQ.
 
A straightforward extension of this work is to include a confining interaction between the fermions. The model should
then be applicable to heavy quarkonia systems. A natural choice for the confining potential in BLFQ is a quadratic confinement.
Such a quadratic potential is motivated by the phenomenological success of the ``soft wall'' AdS/QCD model \cite{brodsky_prl,brodsky_ads}. 
The effective interaction implemented here would then 
be interpreted as providing QCD corrections to the semiclassical approximation provided by the AdS/QCD model.

The ultimate goal, however, is to incorporate the higher Fock sectors into the basis, thereby including dynamical bosons.
This will be required to obtain realistic QCD results without an {\it ad hoc} confining interaction. The main challenge is the
development and implementation of a non-perturbative renormalization scheme, such as the Fock sector dependent scheme of 
Refs. \cite{Karmanov:2008br,Karmanov:2012aj}.
The recent successful renormalization of the free electron problem in BLFQ \cite{Zhao:2014xaa} with a dynamical photon in the basis suggests the possibility
of ``embedding'' the renormalized, physical electron into the positronium system \cite{embed}. Such developments would make BLFQ a 
powerful tool for addressing the {\it ab initio} properties of the hadrons.

\begin{acknowledgments}
We thank S. J. Brodsky, H. Honkanen, D. Chakrabarti and V. A. Karmanov for fruitful discussions.
This work was supported in part by the Department
of Energy under Grant Nos. DE-FG02-87ER40371 and
DESC0008485 (SciDAC-3/NUCLEI) and by the National
Science Foundation under Grant No. PHY-0904782.
A portion of the computational resources
were provided by the National Energy Research Scientific Computing 
Center (NERSC), which is supported by the Office of Science 
of the U.S. Department of Energy under Contract No. DE-AC02-05CH11231.
\end{acknowledgments}

\appendix*

\section{Derivation of Two-Body Effective Interaction}

\label{sec:appendix}

In this Appendix, we present the detailed derivation of the two-body effective interaction, including the necessary integrals. 
Our starting point is the Bloch effective Hamiltonian \cite{dipankar}:
\begin{align}
\Bra{f}H_{\text{eff}}\Ket{i}&=\frac{1}{2}\sum_{n}\Bra{f}{\cal P}H{\cal Q}\Ket{n}\Bra{n}{\cal Q }H{\cal P}\Ket{i}\nonumber\\
&\times\left[\frac{1}{\epsilon_i-\epsilon_n}+
\frac{1}{\epsilon_f-\epsilon_n}\right].
\label{eq:heff}
\end{align}
For this BLFQ application, the $P$ space is the $\ket{e^+e^-}$ sector, while the $Q$ space is the $\ket{e^+e^-\gamma}$ sector.
Also, in BLFQ applications, $H\equiv H_{LC}=P^\mu P_\mu=P^+P^- -\mathbf{P}^2$.
Note that $\mathbf{P}^2$ does not couple the sectors, and so does not contribute to this summation.
Therefore, here we will use $H=P^+P^-$. The sum is over the complete $Q$ space. The notation $\epsilon_i$ denotes the eigenvalue
of the free Hamiltonian $H_0$ for the state $\Ket{i}$ (see Eq. \eqref{h0}):
\begin{equation}
\epsilon_i=\sum_j\frac{\mathbf{p}_j^2+m_j^2}{x_j}, 
\label{eq:kinetic}
\end{equation}
where the sum runs over all particles (of mass $m_j$) in the state $\Ket{i}$.

After adding the counterterm presented in the main text, the effective Hamiltonian takes the form:
\begin{equation}
\Bra{f}H^{\new}_{\text{eff}}\Ket{i}=\sum_{n}\frac{\Bra{f}{\cal P}H{\cal Q}\Ket{n}\Bra{n}{\cal Q}H{\cal P}\Ket{i}}{\frac{1}{2}
\left[\left(\epsilon_i-\epsilon_n\right)+\left(\epsilon_f-\epsilon_n\right)\right]}.
\end{equation}
We will see that this choice, and {\it only} this choice, results in the cancellation of the light-front small-{\it x} divergences. 

\subsection{Sum Over Intermediate States}
The basic interaction of the LFQED Hamiltonian that connects the sectors is the vertex interaction, given by
\begin{equation}
P^-=g\int_{-L}^{L}dx^-d^2\mathbf{x}\overline{\Psi}(x)\gamma_\mu A^\mu(x) \Psi(x),
\end{equation}
evaluated at $x^+=0$. 

The free field mode expansions in BLFQ are 
\begin{align}
\Psi(x)=\sum_\alpha \frac{1}{\sqrt{2L}}&\int \frac{d^2\mathbf{p}}{(2\pi)^2}\left[b_\alpha(\mathbf{p})u_s(p)e^{-ipx}\right.\nonumber\\
&\qquad \left. +d^\dagger_\alpha(\mathbf{p})v_s(p)e^{ipx}\right]
\end{align}
\begin{equation}
A^\mu(x)=\sum_\beta \frac{1}{\sqrt{2Lk^+}}\int \frac{d^2\mathbf{k}}{(2\pi)^2}\left[a_\beta(\mathbf{k})\epsilon^\mu_\lambda(k)e^{-ikx}
+\text{h.c.}\right],
\end{equation}
where the greek subscripts are shorthand for the set of quantum numbers $\alpha=(j,s)$ and $\beta=(j,\lambda)$, 
where $s=\pm\frac{1}{2}$ is the fermion spin and 
$\lambda=\pm1$ is the photon helicity. 

\begin{figure}[t]
\includegraphics[width=3.3 in]{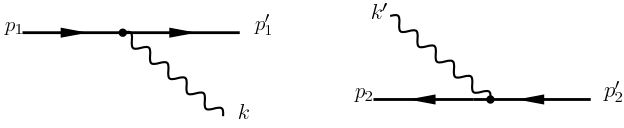}
\caption{Basic Hamiltonian interaction vertices: $P^-_{e \to e\gamma}$ (left) $P^-_{\bar{e}\gamma \to \bar{e}}$ (right). }
\label{fig:vertices}
\end{figure}

Plugging these mode expansions into $P^-$ yields terms of the form $P^-_{e\gamma \to e}\sim b^\dagger b a$, 
$P^-_{e \to e\gamma}\sim b^\dagger b a^\dagger$, $P^-_{\bar{e}\gamma \to \bar{e}}\sim d^\dagger d a$, 
$P^-_{\bar{e} \to \bar{e}\gamma}\sim d^\dagger d a^\dagger$.
These are the only terms which will connect the $\ket{e^+e^-}$ states to the $\ket{e^+e^-\gamma}$ states and survive the operation of 
the projection operators. For example we have the basic interaction vertices shown in Fig. \ref{fig:vertices}, which are given by
\begin{align}
P^-_{e \to e\gamma}&=g\sum_{\alpha_1\alpha_1'\beta}\frac{\theta(p_1^+ -p_1'^+)}{\sqrt{4\pi Kx_k}}\delta_{j_1}^{j_k+j_1'} \nonumber \\
&\times\int \frac{d^2\mathbf{p}}{(2\pi)^2} \frac{d^2\mathbf{p}_1'}{(2\pi)^2} \frac{d^2\mathbf{k}}{(2\pi)^2}
(2\pi)^2\delta^{(2)}(\mathbf{p}_1'+\mathbf{k} -\mathbf{p}_1)\nonumber\\
&\times\bar{u}_{s_1'}(p_1')\gamma_\mu\epsilon^{\mu *}_\lambda(k)u_{s_1}(p_1)b^\dagger_{\alpha_1'}(\mathbf{p}_1')
a^\dagger_\beta(\mathbf{k})b_{\alpha_1}(\mathbf{p}_1)
\label{e->eg}
\end{align}
and
\begin{align}
P^-_{\bar{e}\gamma \to \bar{e}}&=-g\sum_{\alpha_2\alpha_2'\beta'}\frac{\theta(p_2'^+ -p_2^+)}{\sqrt{4\pi Kx_k'}}\delta_{j_2+j_k'}^{j_2'}\nonumber\\
&\times\int \frac{d^2\mathbf{p}_2}{(2\pi)^2} \frac{d^2\mathbf{p}_2'}{(2\pi)^2} \frac{d^2\mathbf{k}'}{(2\pi)^2}
(2\pi)^2\delta^{(2)}(\mathbf{p}_2+\mathbf{k}' -\mathbf{p}_2')\nonumber\\
&\times\bar{v}_{s_2}(p_2)\gamma_\mu\epsilon^\mu_{\lambda'}(k')v_{s_2'}(p_2')
d^\dagger_{\alpha_2'}(\mathbf{p}_2')d_{\alpha_2}(\mathbf{p}_2)a_{\beta'}(\mathbf{k}').
\label{ebarg->ebar}
\end{align}
The overall minus in Eq. \eqref{ebarg->ebar} has arisen from normal ordering, dropping a vacuum bubble.
The theta functions appear since the light-front longitudinal momentum must be positive.

\begin{figure}[t]
\includegraphics[width=3.3 in]{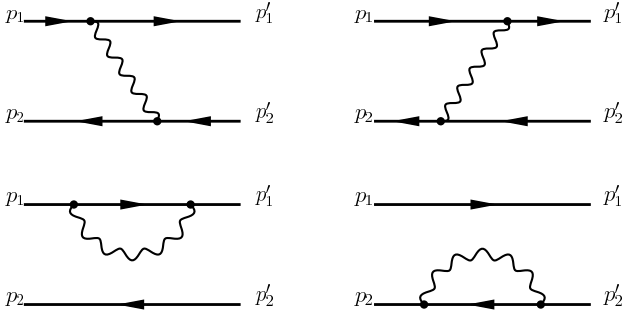}
\caption{Iterated interactions generated in the two-body effective interaction. The photon exchange diagrams correspond to
$(H^{\new}_{\text{eff}})_{e \to e\gamma}$ (left) and $(H^{\new}_{\text{eff}})_{\bar{e} \to \bar{e}\gamma}$
(right). The fermion self energy contributions are neglected in $H^{\new}_{\text{eff}}$. }
\label{fig:iterated}
\end{figure}

In the effective interaction, these basic interaction vertices are stitched together to generate both 
fermion self-energy loops and exchanges of photons
between the electron and positron as shown in Fig. \ref{fig:iterated}.   
From here on, we neglect these fermion self energy terms to focus on the photon exchange terms.
The effective Hamiltonian is then:
\begin{align}
H^{\new}_{\text{eff}}&=\sum_{n}\frac{P^+P^-_{\bar{e}\gamma \to \bar{e}}\Ket{n}\Bra{n}P^+P^-_{e \to e\gamma}}{\frac{1}{2}
\left[\left(\epsilon_i-\epsilon_n\right)+\left(\epsilon_f-\epsilon_n\right)\right]}\nonumber\\
&+\sum_{n}\frac{P^+P^-_{e\gamma \to e}\Ket{n}\Bra{n}P^+P^-_{\bar{e} \to \bar{e}\gamma}}{\frac{1}{2}
\left[\left(\epsilon_i-\epsilon_n\right)+\left(\epsilon_f-\epsilon_n\right)\right]}\nonumber\\
&\equiv (H^{\new}_{\text{eff}})_{e \to e\gamma}+(H^{\new}_{\text{eff}})_{\bar{e} \to \bar{e}\gamma}.
\end{align}

We now turn to a particular time ordering of the photon exchange, namely the one in which the photon is first emitted by the electron
and later absorbed by the positron. This corresponds to the combination $(H^{\new}_{\text{eff}})_{e \to e\gamma}$ described
above. Let us consider the expression for the effective Hamiltonian in BLFQ. 
Instead of doing an infinite summation over the complete ${\cal Q }$ space in the H.O. basis, we perform the summation in momentum space, where 
the sum is an integral and we can 
write
\begin{eqnarray}
&&(H^{\new}_{\text{eff}})_{e \to e\gamma}=\sum_{n}\frac{P^+P^-_{\bar{e}\gamma \to \bar{e}}\Ket{n}\Bra{n}P^+P^-_{e \to e\gamma}}{\frac{1}{2}
\left[\left(\epsilon_i-\epsilon_n\right)+\left(\epsilon_f-\epsilon_n\right)\right]}\nonumber \\
&=&(P^+)^2\sum_{\delta\epsilon\xi}\int\frac{d^2\mathbf{r}}{(2\pi)^2}
\frac{d^2\mathbf{s}}{(2\pi)^2}\frac{d^2\mathbf{t}}{(2\pi)^2}\nonumber \\
&&\times\frac{P^-_{\bar{e}\gamma \to \bar{e}} b^\dagger_\delta(\mathbf{r}) 
d^\dagger_\epsilon(\mathbf{s}) a^\dagger_\xi(\mathbf{t})
\Ket{0}\Bra{0}a_\xi(\mathbf{t})d_\epsilon(\mathbf{s})b_\delta(\mathbf{r})P^-_{e \to e\gamma}}{{\cal D}_{\delta,\epsilon,\xi}(\mathbf{r},\mathbf{s},\mathbf{t})}.
\nonumber\\
\label{eq:heff_new}
\end{eqnarray}
The energy denominator is given by
\begin{align}
&{\cal D}_{\delta,\epsilon,\xi}(\mathbf{r},\mathbf{s},\mathbf{t})=\frac{1}{2}
\left[\left(\epsilon_i-\epsilon_n\right)+\left(\epsilon_f-\epsilon_n\right)\right]=
\frac{\epsilon_i+\epsilon_f}{2}-\epsilon_n\nonumber \\
&=\frac{1}{2}\!\left[\! \left(\!\frac{\mathbf{p}_1^2+m_f^2}{x_1}+\frac{\mathbf{p}_2^2+m_f^2}{x_2}\right)\!\!+\!\!   
\left(\!\frac{\mathbf{p}_1'^2+m_f^2}{x_1'} +\frac{\mathbf{p}_2'^2+m_f^2}{x_2'}\right)  \!\right]\nonumber \\
&-\left(\frac{\mathbf{r}^2+m_f^2}{x_r}+\frac{\mathbf{s}^2+m_f^2}{x_s}+\frac{\mathbf{t}^2+\mu^2}{x_t}\right).
\end{align}
In the final equality we have used the expression for the kinetic energy appropriate for $H=P^+P^-$. For simplicity of notation, below we will
write ${\cal D}_{\delta,\epsilon,\xi}(\mathbf{r},\mathbf{s},\mathbf{t})\equiv{\cal D}(r,s,t)$.

We can now substitute Eqs. \eqref{e->eg} and \eqref{ebarg->ebar} into Eq. \eqref{eq:heff_new}. With the help of the 
canonical commutation relations
\begin{equation}
\left[a_\gamma(\mathbf{k}),a^\dagger_{\gamma'}(\mathbf{k}')\right]_-=(2\pi)^2\delta^{(2)}(\mathbf{k}-\mathbf{k}')\delta_{\gamma}^{\gamma'}
\end{equation}
\begin{align}
\left[b_\alpha(\mathbf{p}),b^\dagger_{\alpha'}(\mathbf{p}')\right]_+
&=(2\pi)^2\delta^{(2)}(\mathbf{p}-\mathbf{p}')\delta_{\alpha}^{\alpha'}\nonumber\\
\left[d_\alpha(\mathbf{p}),d^\dagger_{\alpha'}(\mathbf{p}')\right]_+
&=(2\pi)^2\delta^{(2)}(\mathbf{p}-\mathbf{p}')\delta_{\alpha}^{\alpha'}
\end{align}
it is straightforward to obtain ($\alpha=\frac{g^2}{4\pi}$):
\begin{align}
&(H^{\new}_{\text{eff}})_{e \to e\gamma}=-\frac{\alpha}{K}\sum_{\alpha_1\alpha_1'\alpha_2\alpha_2'}\delta^{j_1'+j_2'}_{j_1+j_2}
\theta(p_1^+-p_1'^+)\nonumber\\
&\times\int \frac{d^2\mathbf{p}_1}{(2\pi)^2} \frac{d^2\mathbf{p}_1'}{(2\pi)^2} \frac{d^2\mathbf{p}_2}{(2\pi)^2} \frac{d^2\mathbf{p}_2'}{(2\pi)^2}
\frac{(2\pi)^2\delta^{(2)}(\mathbf{p}_1\!+\!\mathbf{p}_2\!-\!\mathbf{p}_1'\!-\!\mathbf{p}_2')}{(x_1-x_1')
{\cal D}(p_1',p_2,p_1-p_1')} \nonumber \\
&\times(P^+)^2\bar{u}_{s_1'}(p_1')\gamma_\mu u_{s_1}(p_1)\bar{v}_{s_2}(p_2)\gamma_\nu v_{s_2'}(p_2')\nonumber\\
&\times\!\!\left(\!\sum_\lambda \epsilon^\mu_
\lambda(k)\epsilon^{\nu *}_\lambda(k)\!\!\right)
b_{\alpha_1'}^\dagger(\mathbf{p}_1')d^\dagger_{\alpha_2'}(\mathbf{p}_2')\Ket{0}\Bra{0}d_{\alpha_2}(\mathbf{p}_2)b_{\alpha_1}(\mathbf{p}_1),
\nonumber\\
\label{heff_e->eg}
\end{align}
where 
$k^\mu=(k^-,k^+,\mathbf{k})\equiv\left(\frac{(\mathbf{p}_1-\mathbf{p}_1')^2}{p_1^+-p_1'^+},p_1^+-p_1'^+,\mathbf{p}_1-\mathbf{p}_1'\right)$. 
Note that $k^\mu\ne(p_1-p_1')^\mu$ due to the minus component. (Light-front energy is not conserved.)
Note that in momentum space the state of the photon is completely determined by the external legs, except for the helicity. Thus there 
remains a sum over helicity states. In the HO basis there would have been an infinite sum over the quantum number $n$ of the 
oscillator.

Similarly, the other time ordering (corresponding to photon emission by the positron and absorption by the electron) gives
\begin{align}
&(H^{\new}_{\text{eff}})_{\bar{e} \to \bar{e}\gamma}=-\frac{\alpha}{K}\sum_{\alpha_1\alpha_1'\alpha_2\alpha_2'}\delta^{j_1'+j_2'}_{j_1+j_2}
\theta(p_2^+-p_2'^+)\nonumber \\
&\times\int \frac{d^2\mathbf{p}_1}{(2\pi)^2} \frac{d^2\mathbf{p}_1'}{(2\pi)^2} \frac{d^2\mathbf{p}_2}{(2\pi)^2} \frac{d^2\mathbf{p}_2'}{(2\pi)^2}
\frac{(2\pi)^2\delta^{(2)}(\mathbf{p}_1\!+\!\mathbf{p}_2\!-\!\mathbf{p}_1'\!-\!\mathbf{p}_2')}{(x_2-x_2')
{\cal D}(p_2',p_1,p_2-p_2')} \nonumber \\
&\times(P^+)^2\bar{u}_{s_1'}(p_1')\gamma_\mu u_{s_1}(p_1)\bar{v}_{s_2}(p_2)\gamma_\nu v_{s_2'}(p_2')\nonumber\\
&\times\!\!\left(\!\sum_\lambda \epsilon^\mu_
\lambda(k)\epsilon^{\nu *}_\lambda(k)\!\!\right)
b_{\alpha_1'}^\dagger(\mathbf{p}_1')d^\dagger_{\alpha_2'}(\mathbf{p}_2')\Ket{0}\Bra{0}d_{\alpha_2}(\mathbf{p}_2)b_{\alpha_1}(\mathbf{p}_1).
\nonumber\\
\label{heff_ebar->ebarg}
\end{align}
Now, in light-cone gauge, the polarization sum is given by
\begin{equation}
\sum_\lambda \epsilon^\mu_\lambda(k)\epsilon^{\nu *}_\lambda(k)=-g^{\mu\nu}+\frac{\eta^\mu k^\nu+\eta^\nu k^\mu}{k^+},
\end{equation}
where $\eta^\mu=(\eta^-,\eta^+,\boldsymbol{\eta})=(2,0,\mathbf{0})$ is a unit vector in the light-front ``$+$'' direction ($k^\mu\eta_\mu=k^+$).
The delta function requires $p_2^+-p_2'^+=-(p_1^+-p_1'^+)$ (similarly for momentum fractions). Also, one can easily show that
${\cal D}(p_2',p_1,p_2-p_2')=-{\cal D}(p_1',p_2,p_1-p_1')$. Therefore, the two time orderings can be combined to obtain our result:
\begin{align}
&H^{\new}_{\text{eff}}=-\frac{\alpha}{K}\sum_{\alpha_1\alpha_1'\alpha_2\alpha_2'}\delta^{j_1'+j_2'}_{j_1+j_2}
\int \frac{d^2\mathbf{p}_1}{(2\pi)^2} \frac{d^2\mathbf{p}_1'}{(2\pi)^2} \frac{d^2\mathbf{p}_2}{(2\pi)^2} \frac{d^2\mathbf{p}_2'}{(2\pi)^2}\nonumber\\
&\times\frac{(2\pi)^2\delta^{(2)}(\mathbf{p}_1+\mathbf{p}_2-\mathbf{p}_1'-\mathbf{p}_2')}{(x_1-x_1'){\cal D}(p_1',p_2,p_1-p_1')}
\left(-g^{\mu\nu}+\frac{\eta^\mu k^\nu+\eta^\nu k^\mu}{k^+}\right)\nonumber\\
&\times(P^+)^2\bar{u}_{s_1'}(p_1')\gamma_\mu u_{s_1}(p_1)\bar{v}_{s_2}(p_2)\gamma_\nu v_{s_2'}(p_2')\nonumber \\
&\times b_{\alpha_1'}^\dagger(\mathbf{p}_1')d^\dagger_{\alpha_2'}(\mathbf{p}_2')d_{\alpha_2}(\mathbf{p}_2)b_{\alpha_1}(\mathbf{p}_1).
\label{result}
\end{align}
Note that the $\Ket{0}\Bra{0}$ present in Eqs. \eqref{heff_e->eg} and \eqref{heff_ebar->ebarg} is redundant within the $\Ket{e^+e^-}$ Fock sector, 
and has therefore been dropped from Eq. \eqref{result}.

\subsection{Cancelling the Instantaneous Photon Exchange Term}
\begin{figure}[t]
\includegraphics[width=1.6 in]{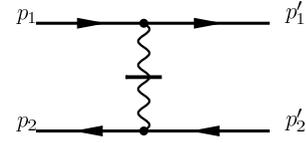}
\caption{Instantaneous photon exchange interaction, $P^-_{\text{inst}}$. The tick mark on the photon denotes instantaneous exchange 
(in light-front time).}
\label{fig:inst}
\end{figure}

The instantaneous photon exchange term in LFQED is given by
\begin{equation}
P^-_{\text{inst}}=\frac{g^2}{2}\int_{-L}^Ldx^-d^2\mathbf{x}\overline{\Psi}(x)\gamma^+\Psi(x)\frac{1}{(i\partial^+)^2}\overline{\Psi}(x)\gamma^+\Psi(x),
\end{equation}
evaluated at $x^+=0$. This is shown diagrammatically in Fig. \ref{fig:inst}. 
Again, we substitute in the free field mode expansions and expand. The term of interest is:
\begin{align}
&H_{\text{inst}}=P^+P^-_{\text{inst}}=-\frac{4\alpha}{K}\sum_{\alpha_1\alpha_1'\alpha_2\alpha_2'}\delta^{j_1'+j_2'}_{j_1+j_2}
\frac{\delta_{s_1}^{s_1'}\delta_{s_2}^{s_2'}}{(x_1-x_1')^2} \nonumber \\
&\times\int \frac{d^2\mathbf{p}_1}{(2\pi)^2} \frac{d^2\mathbf{p}_1'}{(2\pi)^2} \frac{d^2\mathbf{p}_2}{(2\pi)^2} \frac{d^2\mathbf{p}_2'}{(2\pi)^2}
(2\pi)^2\delta^{(2)}(\mathbf{p}_1\!+\!\mathbf{p}_2\!-\!\mathbf{p}_1'\!-\!\mathbf{p}_2')\nonumber \\
&\times b_{\alpha_1'}^\dagger(\mathbf{p}_1')d^\dagger_{\alpha_2'}(\mathbf{p}_2')d_{\alpha_2}(\mathbf{p}_2)b_{\alpha_1}(\mathbf{p}_1).
\label{eq:inst}
\end{align}

We will now show that the second term in the parentheses of $H^{\new}_{\text{eff}}$ (Eq. \eqref{result}) exactly cancels $H_{\text{inst}}$.
Comparing, we see that cancellation will be achieved if 
\begin{align}
&A=\frac{(P^+)^2}{4(x_1-x_1'){\cal D}(p_1',p_2,p_1-p_1')}\frac{\eta^\mu k^\nu+\eta^\nu k^\mu}{k^+}\nonumber\\
&\times\bar{u}_{s_1'}(p_1')\gamma_\mu u_{s_1}(p_1)\bar{v}_{s_2}(p_2)\gamma_\nu v_{s_2'}(p_2')=-\frac{\delta_{s_1}^{s_1'}\delta_{s_2}^{s_2'}}{(x_1-x_1')^2}.
\end{align}
This is indeed the case, but before we proceed we need a few results. Consider the four-vector $l_e^\mu\equiv(k+p_1'-p_1)^\mu$. By the 
definition of $k^\mu$ (just after Eq. \ref{heff_e->eg}), we have $l_e^+=0$ and $l_e^\perp=0$. Thus $l_e^\mu\sim\eta^\mu$.
In fact, if we define $Q\equiv l_e^-$, then we have $l_e^\mu=\frac{Q}{2}\eta^\mu$ (recall $\eta^-=2$). Combining this with the 
definition of $l_e^\mu$, we see that
\begin{equation}
k^\mu=\frac{Q}{2}\eta^\mu+(p_1-p_1')^\mu,
\label{Q}
\end{equation}
where $Q=l_e^-=k^-+p_1'^- -p_1^-$. Similarly, starting from the vector $l_{\bar{e}}^\mu \equiv(k+p_2-p_2')^\mu$, we can show that
\begin{equation}
k^\mu=\frac{\bar{Q}}{2}\eta^\mu+(p_2'-p_2)^\mu,
\label{barQ}
\end{equation}
where $\bar{Q}=l_{\bar{e}}^-=k^- +p_2^- -p_2'^-$. Using this information, we see that
\begin{align}
\bar{u}_{s_1'}(p_1')\gamma_\mu k^\mu u_{s_1}(p_1)&=\frac{Q}{2}\bar{u}_{s_1'}(p_1')\gamma^+ u_{s_1}(p_1)=\frac{Q}{2}(2\delta_{s_1}^{s_1'}) \nonumber\\
\bar{v}_{s_2}(p_2)\gamma_\nu k^\nu v_{s_2'}(p_2')&=\frac{\bar{Q}}{2}\bar{v}_{s_2}(p_2)\gamma^+ v_{s_2'}(p_2')=\frac{\bar{Q}}{2}(2\delta_{s_2}^{s_2'}).
\end{align}
Use has been made of the Dirac spinor identity $(p-p')^\mu \bar{u}_{s'}(p')\gamma_\mu u_{s}(p)=0$ (similarly for the positron) 
and $\gamma_\mu \eta^\mu = \gamma^+$. 

We are now ready to consider
\begin{align}
A&=\frac{(P^+)^2}{4(x_1-x_1'){\cal D}(p_1',p_2,p_1-p_1')}\frac{\eta^\mu k^\nu+\eta^\nu k^\mu}{k^+}\nonumber\\
&\times\bar{u}_{s_1'}(p_1')\gamma_\mu u_{s_1}(p_1)\bar{v}_{s_2}(p_2)\gamma_\nu v_{s_2'}(p_2')\nonumber\\
&=\frac{(2\delta_{s_1}^{s_1'})(2\delta_{s_2}^{s_2'})}{4(x_1-x_1')^2}\left[\frac{\frac{Q+\bar{Q}}{2}P^+}{{\cal D}(p_1',p_2,p_1-p_1')}\right].
\end{align}
By writing out the expressions for $Q$,$\bar{Q}$ and ${\cal D}$, it is readily seen that that the factor in brackets is $-1$. Therefore
\begin{equation}
A=-\frac{\delta_{s_1}^{s_1'}\delta_{s_2}^{s_2'}}{(x_1-x_1')^2}
\end{equation}
as required for the cancellation of the instantaneous term. Therefore we obtain:
\begin{align}
&H^{\new}_{\text{eff}}+H_{\text{inst}}=\frac{\alpha}{K}\sum_{\alpha_1\alpha_1'\alpha_2\alpha_2'}\delta^{j_1'+j_2'}_{j_1+j_2}\nonumber\\
&\times\int \frac{d^2\mathbf{p}_1}{(2\pi)^2} \frac{d^2\mathbf{p}_1'}{(2\pi)^2} \frac{d^2\mathbf{p}_2}{(2\pi)^2} \frac{d^2\mathbf{p}_2'}{(2\pi)^2}
\frac{(2\pi)^2\delta^{(2)}(\mathbf{p}_1\!+\!\mathbf{p}_2\!-\!\mathbf{p}_1'\!-\!\mathbf{p}_2')}{(x_1-x_1'){\cal D}(p_1',p_2,p_1-p_1')}\nonumber\\
&\times (P^+)^2\bar{u}_{s_1'}(p_1')\gamma_\mu u_{s_1}(p_1)\bar{v}_{s_2}(p_2)\gamma^\mu v_{s_2'}(p_2')\nonumber \\
&\times b_{\alpha_1'}^\dagger(\mathbf{p}_1')d^\dagger_{\alpha_2'}(\mathbf{p}_2')d_{\alpha_2}(\mathbf{p}_2)b_{\alpha_1}(\mathbf{p}_1),
\label{result2}
\end{align}
where we have contracted the $g^{\mu\nu}$ term.

\subsection{Translation Back to Harmonic Oscillator Basis}
To translate the result \eqref{result2} to the HO basis, we use
\begin{equation}
b_\alpha(\mathbf{p})=\sum_{nm}b_{\bar{\alpha}}\Psi_n^m(\mathbf{p}),
\end{equation}
where $\bar{\alpha}$ is now shorthand for the set of quantum numbers $(j,s,n,m)$. We therefore obtain the result
\begin{align}
&H^{\new}_{\text{eff}}+H_{\text{inst}}=\frac{\alpha}{K}\sum_{\bar{\alpha}_1\bar{\alpha}_1'\bar{\alpha}_2\bar{\alpha}_2'}
\delta^{j_1'+j_2'}_{j_1+j_2}
b_{\bar{\alpha}_1'}^\dagger d^\dagger_{\bar{\alpha}_2'}d_{\bar{\alpha}_2}b_{\bar{\alpha}_1}\nonumber\\
&\times\int \frac{d^2\mathbf{p}_1}{(2\pi)^2} \frac{d^2\mathbf{p}_1'}{(2\pi)^2} \frac{d^2\mathbf{p}_2}{(2\pi)^2} \frac{d^2\mathbf{p}_2'}{(2\pi)^2}
\frac{(2\pi)^2\delta^{(2)}(\mathbf{p}_1\!+\!\mathbf{p}_2\!-\!\mathbf{p}_1'\!-\!\mathbf{p}_2')}{(x_1-x_1'){\cal D}(p_1',p_2,p_1-p_1')}\nonumber\\
&\times\Psi_{n_1}^{m_1}(\mathbf{p}_1)\Psi_{n_2}^{m_2}(\mathbf{p}_2)\Psi_{n_1'}^{m_1'\ast}(\mathbf{p}_1')\Psi_{n_2'}^{m_2'\ast}(\mathbf{p}_2')\nonumber \\
&\times (P^+)^2\bar{u}_{s_1'}(p_1')\gamma_\mu u_{s_1}(p_1)\bar{v}_{s_2}(p_2)\gamma^\mu v_{s_2'}(p_2').
\label{result3}
\end{align}
The spinor part 
$(P^+)^2\bar{u}_{s_1'}(p_1')\gamma_\mu u_{s_1}(p_1)\bar{v}_{s_2}(p_2)\gamma^\mu v_{s_2'}(p_2')$
contains 16 different spin combinations. These are enumerated in Table \ref{tab:spinor}.

To translate \eqref{result3} into the longitudinal momentum weighted coordinates, we make the substitutions $\mathbf{p}\to\sqrt{x}\mathbf{q}$,
$\int d^2\mathbf{p}\to x\int d^2\mathbf{q}$, $b_{\bar{\alpha}}\to b_{\bar{\alpha}}/\sqrt{x}$ and $\Psi_n^m({\mathbf{p}})\to \Psi_n^m({\mathbf{q}})$.  
Eq. \eqref{result3}  now reads
\begin{align}
&H^{\new}_{\text{eff}}+H_{\text{inst}}=\frac{\alpha}{K}\sum_{\bar{\alpha}_1\bar{\alpha}_1'\bar{\alpha}_2\bar{\alpha}_2'}
\delta^{j_1'+j_2'}_{j_1+j_2}
b_{\bar{\alpha}_1'}^\dagger d^\dagger_{\bar{\alpha}_2'}d_{\bar{\alpha}_2}b_{\bar{\alpha}_1}
\nonumber\\
&\times\sqrt{x_1x_2x_1'x_2'}\int \frac{d^2\mathbf{q}_1}{(2\pi)^2} \frac{d^2\mathbf{q}_1'}{(2\pi)^2} \frac{d^2\mathbf{q}_2}{(2\pi)^2} 
\frac{d^2\mathbf{q}_2'}{(2\pi)^2}\nonumber \\
&\times\frac{(2\pi)^2\delta^{(2)}(\sqrt{x_1}\mathbf{q}_1+\sqrt{x_2}\mathbf{q}_2-\sqrt{x_1'}\mathbf{q}_1'-\sqrt{x_2'}\mathbf{q}_2')}
{(x_1-x_1'){\cal D}}\nonumber\\
&\times\Psi_{n_1}^{m_1}(\mathbf{q}_1)\Psi_{n_2}^{m_2}(\mathbf{q}_2)\Psi_{n_1'}^{m_1'\ast}(\mathbf{q}_1')\Psi_{n_2'}^{m_2'\ast}(\mathbf{q}_2')\nonumber \\
&\times S_{\alpha_1,\alpha_2,\alpha_1',\alpha_2'}(\sqrt{x_1}\mathbf{q}_1,\sqrt{x_2}\mathbf{q}_2,\sqrt{x_1'}\mathbf{q}_1',\sqrt{x_2'}\mathbf{q}_2'),
\label{result3.5}
\end{align}
where the energy denominator factor ${\cal D}$ is now given by
\begin{align}
{\cal D}&=\frac{1}{2}\left[\left(\frac{(\sqrt{x_1}\mathbf{q}_1)^2+m_f^2}{x_1}-\frac{(\sqrt{x_1'}\mathbf{q}'_1)^2+m_f^2}{x_1'}\right.\right.\nonumber\\
&\qquad \left. \left. -\frac{(\sqrt{x_1}\mathbf{q}_1-\sqrt{x_1'}\mathbf{q}_1')^2+\mu^2}{x_1-x_1'}\right)
-(1\to2)\right].
\end{align}

\begin{table}[t]
\caption{Spinor part $S_{\alpha_1,\alpha_2,\alpha_1',\alpha_2'}(\mathbf{p}_1,\mathbf{p}_2,\mathbf{p}_1',\mathbf{p}_2')\equiv
(P^+)^2\bar{u}_{s_1'}(p_1')\gamma_\mu u_{s_1}(p_1)\bar{v}_{s_2}(p_2)\gamma^\mu v_{s_2'}(p_2')$.
In this expression, the notation $p_j$ stands for the complex number $p_j=(\mathbf{p}_j)_x+i(\mathbf{p}_j)_y$. Thus $p_j^\ast=(\mathbf{p}_j)_x-i(\mathbf{p}_j)_y$.}
\begin{tabular}{|c|c|c|c|c|}
\hline
$s_1$ & $s_2$ & $s_1'$ & $s_2'$ & $S_{\alpha_1,\alpha_2,\alpha_1',\alpha_2'}(\mathbf{p}_1,\mathbf{p}_2,\mathbf{p}_1',\mathbf{p}_2')$\\
\hline\hline
$+$&$+$&$+$&$+$&$2m_f^2\left(\frac{1}{x_1x_1'}+\frac{1}{x_2x_2'}\right)+2\left(\frac{p_1'^\ast}{x_1'}-\frac{p_2'^\ast}{x_2'}\right)\left(\frac{p_1}{x_1}-\frac{p_2}{x_2}\right)$\\
\hline
$-$&$-$&$-$&$-$&$2m_f^2\left(\frac{1}{x_1x_1'}+\frac{1}{x_2x_2'}\right)+2\left(\frac{p_1^\ast}{x_1}-\frac{p_2^\ast}{x_2}\right)\left(\frac{p_1'}{x_1'}-\frac{p_2'}{x_2'}\right)$\\
\hline
$+$&$-$&$+$&$-$&$2m_f^2\left(\frac{1}{x_1x_1'}+\frac{1}{x_2x_2'}\right)+2\left(\frac{p_1'^\ast}{x_1'}-\frac{p_2^\ast}{x_2}\right)\left(\frac{p_1}{x_1}-\frac{p_2'}{x_2'}\right)$\\
\hline
$-$&$+$&$-$&$+$&$2m_f^2\left(\frac{1}{x_1x_1'}+\frac{1}{x_2x_2'}\right)+2\left(\frac{p_1^\ast}{x_1}-\frac{p_2'^\ast}{x_2'}\right)\left(\frac{p_1'}{x_1'}-\frac{p_2}{x_2}\right)$\\
\hline
$+$&$+$&$+$&$-$&$2m_f\left[\frac{p_2-p_2'}{x_2x_2'}+\left(\frac{1}{x_2}-\frac{1}{x_2'}\right)\frac{p_1}{x_1}\right]$\\
\hline
$-$&$+$&$-$&$-$&$2m_f\left[\frac{p_2-p_2'}{x_2x_2'}+\left(\frac{1}{x_2}-\frac{1}{x_2'}\right)\frac{p_1'}{x_1'}\right]$\\
\hline
$+$&$-$&$+$&$+$&$2m_f\left[\frac{p_2'^\ast-p_2^\ast}{x_2x_2'}+\left(\frac{1}{x_2'}-\frac{1}{x_2}\right)\frac{p_1'^\ast}{x_1'}\right]$\\
\hline
$-$&$-$&$-$&$+$&$2m_f\left[\frac{p_2'^\ast-p_2^\ast}{x_2x_2'}+\left(\frac{1}{x_2'}-\frac{1}{x_2}\right)\frac{p_1^\ast}{x_1}\right]$\\
\hline
$+$&$+$&$-$&$+$&$2m_f\left[\frac{p_1-p_1'}{x_1x_1'}+\left(\frac{1}{x_1}-\frac{1}{x_1'}\right)\frac{p_2}{x_2}\right]$\\
\hline
$+$&$-$&$-$&$-$&$2m_f\left[\frac{p_1-p_1'}{x_1x_1'}+\left(\frac{1}{x_1}-\frac{1}{x_1'}\right)\frac{p_2'}{x_2'}\right]$\\
\hline
$-$&$+$&$+$&$+$&$2m_f\left[\frac{p_1'^\ast-p_1^\ast}{x_1x_1'}+\left(\frac{1}{x_1'}-\frac{1}{x_1}\right)\frac{p_2'^\ast}{x_2'}\right]$\\
\hline
$-$&$-$&$+$&$-$&$2m_f\left[\frac{p_1'^\ast-p_1^\ast}{x_1x_1'}+\left(\frac{1}{x_1'}-\frac{1}{x_1}\right)\frac{p_2^\ast}{x_2}\right]$\\
\hline
$+$&$-$&$-$&$+$&$2m_f^2\left(\frac{1}{x_1}-\frac{1}{x_1'}\right)\left(\frac{1}{x_2}-\frac{1}{x_2'}\right)$\\
\hline
$-$&$+$&$+$&$-$&$2m_f^2\left(\frac{1}{x_1}-\frac{1}{x_1'}\right)\left(\frac{1}{x_2}-\frac{1}{x_2'}\right)$\\
\hline
$+$&$+$&$-$&$-$&$0$\\
\hline
$-$&$-$&$+$&$+$&$0$\\
\hline
\end{tabular}
\label{tab:spinor}
\end{table}

\subsection{The Energy Denominator Integral}
First we note that the spinor part simply adds factors of the type $p_ip_j^\ast$ to the integrations.  We can easily absorb these complex transverse momenta into the wavefunctions using
\begin{align}
&p\Psi_n^m(\mathbf{p})=\nonumber\\
&b
\begin{cases}
\sqrt{n\!+\!|m|\!+\!1}\Psi_n^{m+1}(\mathbf{p})\!-\!\theta(n\!-\!1)\sqrt{n}\Psi_{n-1}^{m+1}(\mathbf{p})\!\!\!\! &; m\geq0\\
\sqrt{n+|m|}\Psi_n^{m+1}(\mathbf{p})-\sqrt{n+1}\Psi_{n+1}^{m+1}(\mathbf{p}) &; m<0
\end{cases}
\end{align}
\begin{align}
&p^\ast\Psi_n^m(\mathbf{p})=\nonumber\\
&b
\begin{cases}
\sqrt{n\!+\!|m|\!+\!1}\Psi_n^{m-1}(\mathbf{p})\!-\!\theta(n\!-\!1)\sqrt{n}\Psi_{n-1}^{m-1}(\mathbf{p})\!\!\!\! &; m\leq0\\
\sqrt{n+|m|}\Psi_n^{m-1}(\mathbf{p})-\sqrt{n+1}\Psi_{n+1}^{m-1}(\mathbf{p}) &; m>0.
\end{cases}
\end{align}
Therefore we need only consider the integral
\begin{align}
&I_{ED}=
\int \frac{d^2\mathbf{q}_1}{(2\pi)^2} \frac{d^2\mathbf{q}_1'}{(2\pi)^2} \frac{d^2\mathbf{q}_2}{(2\pi)^2} \frac{d^2\mathbf{q}_2'}{(2\pi)^2}\nonumber \\
&\times\frac{(2\pi)^2\delta^{(2)}(\sqrt{x_1}\mathbf{q}_1+\sqrt{x_2}\mathbf{q}_2-\sqrt{x_1'}\mathbf{q}_1'-\sqrt{x_2'}\mathbf{q}_2')}
{(x_1-x_1'){\cal D}}\nonumber\\
&\times\Psi_{n_1}^{m_1}(\mathbf{q}_1)\Psi_{n_2}^{m_2}(\mathbf{q}_2)\Psi_{n_1'}^{m_1'\ast}(\mathbf{q}_1')\Psi_{n_2'}^{m_2'\ast}(\mathbf{q}_2').
\label{result4}
\end{align}
This integral, now expressed in single-particle coordinates, can be simplified by transforming to relative coordinates, using the Talmi-Moshinsky (TM)
transform \cite{yang,Talmi}:
\begin{equation}
\Psi_{n_1}^{m_1}(\mathbf{q}_1)\Psi_{n_2}^{m_2}(\mathbf{q}_2)=\sum_{NMnm}{\cal M}_{n_1m_1n_2m_2}^{NMnm}
\Psi_{N}^{M}(\mathbf{Q})\Psi_{n}^{m}(\mathbf{q}).
\end{equation}
The quantities ${\cal M}_{n_1m_1n_2m_2}^{NMnm}$ are known as TM brackets, and the new relative coordinates are
\begin{align}
&\mathbf{q}=\frac{\sqrt{x_2}\mathbf{q}_1-\sqrt{x_1}\mathbf{q}_2}{\sqrt{x_1+x_2}}\nonumber\\
&\mathbf{Q}=\frac{\sqrt{x_1}\mathbf{q}_1+\sqrt{x_2}\mathbf{q}_2}{\sqrt{x_1+x_2}}.
\label{eq:tmvars}
\end{align}

We now perform two separate TM transformations from the variables
$\mathbf{q}_1,\mathbf{q}_2\to\mathbf{Q},\mathbf{q}$ and $\mathbf{q}_1',\mathbf{q}_2'\to\mathbf{Q'},\mathbf{q'}$.
($\mathbf{Q'}$ and $\mathbf{q'}$ are also given by \eqref{eq:tmvars}, with all quantities primed.)
In terms of the relative coordinates, the denominator can be expressed as
\begin{align}
(x_1-x_1')&{\cal D}=-\frac{1}{2(x_1+x_2)}\left[(\sqrt{x_1'x_2}\mathbf{q}-\sqrt{x_1x_2'}\mathbf{q}')^2\right.\nonumber\\
&\qquad\left.+(\sqrt{x_1'x_2}\mathbf{q}’-\sqrt{x_1x_2'}\mathbf{q})^2 
+(x_1+x_2)\Delta\right],
\end{align}
where $\Delta=m_f^2(x_1-x_1')^2\left[\frac{1}{x_1x_1'}+\frac{1}{x_2x_2'}\right]+2\mu^2\geq0$.
Note the denominator is independent of $\mathbf{Q}$ and $\mathbf{Q}'$.

Due to the $x$ dependence of the coordinates $\mathbf{q}_i$ and $\mathbf{Q}_i$, the TM
brackets themselves have $x$ dependence \cite{yang}. In particular, the TM phase \cite{Talmi,yang} $\delta$ is given by
$\tan\delta=\sqrt{x_2/x_1}$ for the $\mathbf{q}_1,\mathbf{q}_2\to\mathbf{Q},\mathbf{q}$ transform and by 
$\tan\delta=\sqrt{x_2'/x_1'}$ for the $\mathbf{q}_1',\mathbf{q}_2'\to\mathbf{Q'},\mathbf{q'}$ transform.
The integral then takes the form:
\begin{align}
I_{ED}&=
\sum_{nmNM}\sum_{n'm'N'M'}{\cal M}_{n_1m_1n_2m_2}^{nmNM}{\cal M}_{n_1'(-m_1')n_2'(-m_2')}^{n'm'N'M'}\nonumber\\
&\times\int \frac{d^2\mathbf{Q}}{(2\pi)^2} \frac{d^2\mathbf{q}}{(2\pi)^2} \frac{d^2\mathbf{Q'}}{(2\pi)^2} \frac{d^2\mathbf{q'}}{(2\pi)^2}\nonumber \\
&\times\frac{(2\pi)^2\delta^{(2)}(\sqrt{x_1+x_2}\mathbf{Q}-\sqrt{x_1'+x_2'}\mathbf{Q'})}{(x_1-x_1'){\cal D}}\nonumber\\
&\times\Psi_{n}^{m}(\mathbf{q})\Psi_{N}^{M}(\mathbf{Q})\Psi_{n'}^{m'}(\mathbf{q'})\Psi_{N'}^{M'}(\mathbf{Q'}).
\label{result5}
\end{align}
Since the denominator is independent of $\mathbf{Q}$ and $\mathbf{Q}'$ we can write
\begin{align}
&I_{ED}\!=\!
\frac{1}{x_1+x_2}\!\!\sum_{nmNM}\!\sum_{n'\!m'\!N'\!M'}\!\!\!\!\!
{\cal M}_{n_1m_1n_2m_2}^{nmNM}{\cal M}_{n_1'(-m_1')n_2'(-m_2')}^{n'm'N'M'}\nonumber\\
&\times\int \frac{d^2\mathbf{Q}}{(2\pi)^2} \Psi_{N}^{M}(\mathbf{Q})\Psi_{N'}^{M'}(\mathbf{Q})
\int\frac{d^2\mathbf{q}}{(2\pi)^2} \frac{d^2\mathbf{q'}}{(2\pi)^2}
\frac{\Psi_{n}^{m}(\mathbf{q})\Psi_{n'}^{m'}(\mathbf{q'})}{{(x_1-x_1'){\cal D}}}
\label{result6}
\end{align}
which simplifies to
\begin{align}
I_{ED}&\!=\!
\frac{1}{x_1+x_2}\!\!\sum_{nmNM}\!\sum_{n'\!m'\!N'\!M'}\!\!\!\!\!
{\cal M}_{n_1m_1n_2m_2}^{nmNM}{\cal M}_{n_1'(-m_1')n_2'(-m_2')}^{n'm'N'M'}\nonumber\\
&\times\delta_{N}^{N'}\delta_{M}^{-M'}\int\frac{d^2\mathbf{q}}{(2\pi)^2} \frac{d^2\mathbf{q'}}{(2\pi)^2}
\frac{\Psi_{n}^{m}(\mathbf{q})\Psi_{n'}^{m'}(\mathbf{q'})}{(x_1-x_1'){\cal D}}\nonumber\\
&=\frac{1}{x_1+x_2}\sum_{nmNM}\sum_{n'm'}{\cal M}_{n_1m_1n_2m_2}^{nmNM}{\cal M}_{n_1'(-m_1')n_2'(-m_2')}^{n'm'N(-M)}\nonumber\\
&\times\int\frac{d^2\mathbf{q}}{(2\pi)^2} \frac{d^2\mathbf{q'}}{(2\pi)^2}
\frac{\Psi_{n}^{m}(\mathbf{q})\Psi_{n'}^{m'}(\mathbf{q'})}{(x_1-x_1'){\cal D}}.
\label{result7}
\end{align}
We will refer to the remaining integral as
\begin{equation}
I_{nmn'm'}\equiv\int\frac{d^2\mathbf{q}}{(2\pi)^2} \frac{d^2\mathbf{q'}}{(2\pi)^2}
\frac{\Psi_{n}^{m}(\mathbf{q})\Psi_{n'}^{m'}(\mathbf{q'})}{(x_1-x_1'){\cal D}}.
\end{equation}
$I_{nmn'm'}$ contains terms like $\mathbf{q}\cdot\mathbf{q}'$ in the denominator and therefore has angular dependence.
The angular dependence of the denominator can be removed by another TM transform using the variables (here
the phase is given by $\tan\delta=1$)
\begin{align}
&\mathbf{p}=\frac{\mathbf{q}-\mathbf{q}'}{\sqrt{2}}\nonumber\\
&\mathbf{P}=\frac{\mathbf{q}+\mathbf{q}'}{\sqrt{2}}.
\end{align}
The integral is given by
\begin{align}
I_{nmn'm'}=-2&\sum_{N''M''n''m''}{\cal M}_{nmn'm'}^{N''M''n''m''}\nonumber\\
&\times\int\frac{d^2\mathbf{P}}{(2\pi)^2} \frac{d^2\mathbf{p}}{(2\pi)^2}
\frac{\Psi_{n''}^{m''}(\mathbf{p})\Psi_{N''}^{M''}(\mathbf{P})}{c\mathbf{P}^2+c'\mathbf{p}^2+\Delta},
\end{align}
where $c=(\sqrt{x_1'x_2}-\sqrt{x_1x_2'})^2/(x_1+x_2)$ and $c'=(\sqrt{x_1'x_2}+\sqrt{x_1x_2'})^2/(x_1+x_2)$. If one now substitutes
in the expressions for the wavefunctions, the angular integrations are trivial and it straightforward to obtain
\begin{align}
I_{nmn'm'}=-&\frac{1}{2\pi}\sum_{N''n''}{\cal M}_{nmn'm'}^{N''0n''0}\nonumber\\
&\times\int_0^\infty dPdp
\frac{e^{-(P+p)/2}L_{N''}(P)L_{n''}(p)}{cP+c'p+\frac{\Delta}{b^2}},
\end{align}
where $P\equiv\mathbf{P}^2/b^2$ and $p\equiv\mathbf{p}^2/b^2$. $L_n(x)=L_n^0(x)$ are the Laguerre polynomials. 
The remaining 2D integral cannot be evaluated in closed form and is done numerically. The complete integral, $I_{ED}$, 
is then given by
\begin{align}
I_{ED}=-\frac{1}{2\pi}&\frac{1}{x_1+x_2}\sum_{nmNM}\sum_{n'm'}\sum_{N''n''}{\cal M}_{n_1m_1n_2m_2}^{nmNM}\nonumber\\
&\times {\cal M}_{n_1'(-m_1')n_2'(-m_2')}^{n'm'N(-M)}
{\cal M}_{nmn'm'}^{N''0n''0}\nonumber\\
&\times\int_0^\infty dPdp
\frac{e^{-(P+p)/2}L_{N''}(P)L_{n''}(p)}{cP+c'p+\frac{\Delta}{b^2}}.
\label{result8}
\end{align}
The eightfold sum can be reduced significantly to three folds. Each TM transform comes with two Kronecker deltas:
\begin{equation}
{\cal M}_{n_1m_1n_2m_2}^{nmNM}\sim\delta_{2n_1+|m_1|+2n_2+|m_2|}^{2n+|m|+2N+|M|}\delta_{m_1+m_2}^{m+M}.
\end{equation}
One of the six
Kronecker deltas is left over as an angular momentum conserving Kronecker delta out front, while the other five can be used to reduce
the sum down to three folds. The final result is:
\begin{align}
I_{ED}&=-\frac{1}{2\pi}\frac{1}{x_1+x_2}\delta_{m_1+m_2}^{m_1'+m_2'}
\sum_{mNN''}{\cal M}_{n_1m_1n_2m_2}^{N,\lambda-m,\mu-N,m}\nonumber\\
&\times {\cal M}_{n_1',m_1',n_2',m_2'}^{N,\lambda-m,\nu-N,m}
{\cal M}_{\mu-N,m,\nu-N,-m}^{N'',0,\mu+\nu-2N+|m|-N'',0}\nonumber\\
&\times\int_0^\infty dPdp
\frac{e^{-(P+p)/2}L_{N''}(P)L_{\mu+\nu-2N+|m|-N''}(p)}{cP+c'p+\frac{\Delta}{b^2}},
\label{result9}
\end{align}
where $\mu=\mu(m)\equiv n_1+n_2+(|m_1|+|m_2|-|m|-|\lambda-m|)/2$, $\nu=\nu(m)\equiv n_1'+n_2'+(|m_1'|+|m_2'|-|m|-|\lambda-m|)/2$ 
and $\lambda=m_1+m_2$.

\end{document}